%% file: paper_05_01_25.tex
\documentclass[eqsecnum]{elsart}

\usepackage{amssymb}
\usepackage{amsmath}
\usepackage[dvips]{graphicx}
\usepackage{dcolumn}
\usepackage{bm, bbm}
\usepackage{curves}
\usepackage{epic}

\usepackage{epsf, times}
\usepackage{subfigure}
\usepackage{psfrag} 

\usepackage{natbib}
\usepackage{cmmib57}
\usepackage{euscript}

\def\openone{\leavevmode\hbox{\small1\kern-4.2pt\normalsize1}}

\begin{document}

\begin{frontmatter}

\title {From quantum mechanics to classical statistical physics: 
        generalized Rokhsar-Kivelson Hamiltonians and the ``Stochastic 
        Matrix Form'' decomposition}

\author[addr1]{Claudio Castelnovo\corauthref{cor}\thanksref{CC and CC}}
\corauth[cor]{Corresponding author. Fax: +1 617 353 9393} 
\ead{castel@buphy.bu.edu}
\author[addr1]{Claudio Chamon\thanksref{CC and CC}}
\author[addr2]{Christopher Mudry}
\author[addr3]{Pierre Pujol}

\address[addr1]{Physics Department, Boston University, Boston, MA 02215, USA}
\address[addr2]{Paul Scherrer Institut, CH-5232 Villigen PSI, Switzerland}
\address[addr3]{Laboratoire de Physique, {\'E}cole Normale Sup{\'e}rieure, 
                46 All{\'e}e d'Italie, 69364 Lyon Cedex 07, France}

\thanks[CC and CC]
{Supported in part by the NSF Grants DMR-0305482 and DMR-0403997.} 

\begin{abstract}
Quantum Hamiltonians that are fine-tuned to their so-called
Rokhsar-Kivelson (RK) points, first presented in the context of 
quantum dimer models, are defined by their representations in preferred 
bases in which their ground state wave functions are intimately related to 
the partition functions of combinatorial problems of classical statistical 
physics. 
We show that all the known examples of quantum Hamiltonians, when fine-tuned
to their RK points, belong to a larger class of real, symmetric, and 
irreducible matrices that admit what we dub a 
\textit{Stochastic Matrix Form} (SMF) decomposition. 
Matrices that are SMF decomposable are shown to be in one-to-one 
correspondence with stochastic classical systems described by a 
Master equation of the matrix type, hence their name. 
It then follows that the 
equilibrium partition function of the stochastic classical system 
partly controls the zero-temperature quantum phase diagram, 
while the relaxation rates of the stochastic classical system 
coincide with the excitation spectrum of the quantum problem. 
Given a generic quantum Hamiltonian construed as an abstract operator 
defined on some Hilbert space, we prove that there exists a continuous 
manifold of bases in which the representation of the quantum Hamiltonian 
is SMF decomposable, i.e., there is a (continuous) manifold of 
distinct stochastic classical systems related to the same quantum problem. 
Finally, we illustrate with three examples of Hamiltonians fine-tuned
to their RK points, 
the triangular quantum dimer model, 
the quantum eight-vertex model, 
and the quantum three-coloring model on the honeycomb lattice, 
how they can be understood within our framework, and how this allows for 
immediate generalizations, e.g., by adding non-trivial interactions to 
these models. 
\end{abstract}


\end{frontmatter}

\section{Introduction}
\label{sec: Introduction}

The combinatorial problem of counting how many ways there are 
to pack dimers on some given lattice is of relevance to chemists, physicists, 
and mathematicians~\cite{Fowler37,Kasteleyn61,Fisher61,Thurston90}. 
A deep connection between the statistical physics 
of closely-packed, hard-core dimers on two-dimensional lattices, and the 
critical behavior of the two-dimensional Ising model was established in the 
early 60's~\cite{Kasteleyn61,Fisher61}. 
Soon after the discovery of high-temperature superconductivity, a quantum 
version of the classical hard-core dimer problem on the square lattice 
was proposed by Kivelson, Rokhsar, and Sethna as an effective low energy 
theory for a doped Mott insulator%
~\cite{Kivelson87,Rokhsar88,Sachdev89,Ioffe89,Fradkin90,Levitov90,Leung96}. 

This example of a square lattice quantum dimer model 
is interesting and unusual in several ways. 
(1)~There is a one-to-one correspondence between the (dimer) 
basis that spans the underlying Hilbert space 
and the configuration space of the combinatorial problem. 
(2)~The quantum Hamiltonian is the sum over local Hermitian operators, 
each of which encodes the competition between 
a potential energy that favors a local ordering of dimers and a kinetic 
energy term that favors a local quantum superposition of dimers, 
i.e., a local quantum liquid state. 
(3)~For a special value of the ratio between the characteristic potential 
and kinetic energies called the Rokhsar-Kivelson (RK) point, the local 
Hermitian operators entering the Hamiltonian are positive semidefinite 
and the ground state (GS) is the equal-weight superposition of all dimer 
states, i.e., the normalization of the GS is nothing but 
the number of ways to closely-pack hard-core dimers on the square 
lattice~\cite{Rokhsar88}. 
Remarkably, this GS is a very peculiar liquid state since it is 
critical according to the results of Kasteleyn on the classical square 
lattice dimer model~\cite{ Kasteleyn61}. 
(4)~It was realized by Henley that the excitation spectrum of the 
quantum square lattice dimer model at the RK point is identical to the 
spectrum of relaxation rates of the classical square lattice dimer model 
out of thermal equilibrium when equipped with a properly chosen Monte Carlo 
dynamics~\cite{Henley97}. 

Building on the close interplay between the classical and quantum square 
lattice dimer models, Moessner and Sondhi extended the triangular lattice 
classical dimer model to a quantum one and showed that its RK point realizes 
an incompressible spin liquid state characterized by topological 
quantum order and robust to small perturbations~\cite{Moessner01}. 
The triangular lattice quantum dimer model provides a theoretical playground
in which concepts related to fractional quantum numbers, emergent gauge 
symmetries, and deconfinement of gauge quantum numbers take a precise 
form~\cite{Wen04}. 
This work triggered the study of classical and quantum dimer models 
on a variety of lattices in two and three dimensions, 
all quantum ones being characterized by RK points, but 
differing on the presence or not of a two-sublattice structure%
~\cite{Chandra01,Moessner01b,Ioselevich02,Fendley02,Misguich02,%
Krauth03,Sondhi03,Huse03,Moessner03,Hermele04,Fradkin04}. 

It was noted by Ardonne, Fendley, and Fradkin that quantum Hamiltonians
that can be fine-tuned to an RK point 
are not exclusively built from classical dimer 
models and that the RK point can be extended to a line, or, more generally,
to a higher-dimensional region of parameter space~\cite{Ardonne04}. 
More specifically, quantum six- and eight-vertex models 
as well as quantum Hamiltonians not anymore related to combinatorial 
statistical physics were shown to possess a manifold of
RK points~\cite{Ardonne04}. 
Around the same time, Henley pointed out that any classical system endowed 
with linear dynamics in time (through a Master equation of the matrix type) 
can be used to construct a quantum Hamiltonian at an RK point~\cite{Henley04}. 

In spite of the vast literature on the properties of
quantum Hamiltonians fine-tuned to their RK points,
there are still some fundamental questions that remain unanswered. 
What are the generic properties of a quantum Hamiltonian fine-tuned
to an RK point, 
and can they be used to define a more general class of problems? 
Under what conditions can one fine-tune the coupling constants of 
a quantum Hamiltonian to reach an RK point? 
Are all quantum Hamiltonians in this general class related to 
classical statistical systems and if so what is the precise nature 
(static versus dynamic) of this relationship? 
Given that the very notion of an RK point is basis-dependent, what is
the fate of an RK point under a change of basis? 
In this paper we will address and answer these questions, 
thus providing insights into the remarkable relationship between 
quantum systems fine-tuned to their RK points and stochastic classical systems 
described by a Master equation of the matrix type. 

Common to all known quantum Hamiltonians when fine-tuned
to their RK points are the following properties: 
\begin{enumerate}
\item[(I)]~The underlying Hilbert $\mathcal{H}$ space is separable, 
i.e., the set $\mathcal{S}=\{\mathcal{C}\}$ of indices $\mathcal{C}$ 
labeling a basis 
$\mathcal{B}=\{|\mathcal{C}\rangle, \; \mathcal{C}\in\mathcal{S}\}$  
of $\mathcal{H}$ 
is countable.
\item[(II)]~The quantum Hamiltonian 
can be decomposed into the sum of positive-semidefinite Hermitian operators, 
each of which being equal to its square after proper rescaling, 
weighted by positive coefficients. 
In particular, when written in the preferred basis, these operators are 
represented by $2\times2$ matrices acting solely on a two-dimensional 
subspace of the Hilbert space $\mathcal{H}$. 
Any of these operators encodes the competition between a potential energy 
diagonal in the preferred basis and a kinetic energy term that 
favors a pairwise quantum superposition of basis states. 
\item[(III)]~A GS of the quantum Hamiltonian can be found by demanding 
that it is annihilated by each and every Hermitian operator defined in~(II). 
As we shall see, the normalization factor in the GS wavefunction can be 
interpreted as the partition function of a specific classical system defined 
over the phase space $\mathcal{S}$. 
\end{enumerate}

We shall first construct the most general representation of 
a quantum Hamiltonian,
given a preferred basis 
$\mathcal{B}=\{|\mathcal{C}\rangle, \; \mathcal{C}\in\mathcal{S}\}$, 
satisfying conditions~(I-III), and we will show that it has a 
parametrical form that we shall dub the 
\textit{Stochastic Matrix Form} (SMF) decomposition.
The set of all matrices that are SMF decomposable
include all known examples of quantum Hamiltonians fine-tuned to 
their RK points but are not limited to these.
We shall show that conditions~(II) and~(III)  
are represented by the positive-semidefinite decomposition 
conditions~[Eq.~(\ref{eq: condition for positive definite})] and by the 
integrability conditions~[Eq.~(\ref{eq: integrability conds})], respectively. 
In the preferred basis, the positive-semidefinite decomposition conditions 
have a \textit{local} character, while the integrability conditions have a 
\textit{global} character. 
Both set of conditions are automatically satisfied when quantum Hamiltonians 
fine-tuned to an RK point are constructed starting from a 
classical system as in Refs.~\cite{Ardonne04} and~\cite{Henley04}, say. 
Whereas the importance of the local positive-semidefinite decomposition 
conditions has long been implicitly known, the fact that they do not 
necessarily imply the global integrability conditions has been overlooked 
so far. The global integrability conditions play a crucial role when 
the SMF decomposition of a quantum Hamiltonian represented in 
the preferred basis is to be achieved by fine-tuning of its 
coupling constants. 

In answering the third question, 
whether all representations of quantum Hamiltonians 
that are SMF decomposable are related to classical statistical systems 
and if so what is the precise nature of this relationship, 
we will first show that it is always possible to extract from the GS 
wavefunction of a quantum Hamiltonian that is SMF decomposable, 
when written in the preferred basis, 
the partition function of a classical system in thermal equilibrium. 
It then follows that the equal-time correlation functions (of operators 
diagonal in the preferred basis) in this GS can 
be interpreted as equal-time correlation functions in a classical system 
in thermal equilibrium. In that sense, 
the quantum phase diagram of a quantum Hamiltonian 
that is SMF decomposable when written in the preferred basis 
\textit{contains} the thermal phase diagram of a classical system 
in thermal equilibrium. This classical system needs not be a pure 
combinatorial problem, nor one equipped with some chemical potential. 
Instead, it is generically described by a partition function evaluated at 
the inverse temperature $\beta$ and given by a sum over Boltzmann weights 
with configuration energies $E^{\ }_{\mathcal{C}}$ for all 
$\mathcal{C}\in\mathcal{S}$. Moreover, 
as a direct consequence of properties~(II-III), we will show that the 
matrix elements of the quantum Hamiltonian in the preferred basis 
$\mathcal{B}=\{|\mathcal{C}\rangle, \; \mathcal{C}\in\mathcal{S}\}$ 
can be naturally interpreted, via a similarity 
transformation, as transition rates of a Master equation of the matrix type 
for the stochastic evolution of the corresponding classical system
with configuration space $\mathcal{S}$. 
These transition rates satisfy detailed balance precisely with respect to 
the equilibrium probability distribution induced by the GS wavefunction. 
Consequently, the excitation spectrum of any quantum problem
that admits an SMF representation in the preferred basis 
coincides with the relaxation rates of a stochastic classical system 
and, as was observed by Henley~\cite{Henley04}, 
quantum correlation functions at unequal imaginary times in the
GS for operators diagonal in the basis 
$\mathcal{B}=\{|\mathcal{C}\rangle, \; \mathcal{C}\in\mathcal{S}\}$ 
are identical to classical correlation functions at unequal times. 
We will show that 
quantum Hamiltonians that admit an SMF representation in the preferred basis 
are parametrized by the (independent) positive coupling constants 
$w^{\ }_{\mathcal{C},\mathcal{C}'}$ 
and by the (independent) coupling constants 
$E^{\ }_{\mathcal{C}}$. Whereas the coupling constants 
$E^{\ }_{\mathcal{C}}$ determine the GS wavefunction, 
the non-vanishing coupling constants $w^{\ }_{\mathcal{C},\mathcal{C}'}$ 
solely determine the excitation spectrum above the GS wavefunction. 
Remarkably, the excitation spectrum can be manipulated 
by fine-tuning the $w^{\ }_{\mathcal{C},\mathcal{C}'}$ 
without affecting the nature of the GS wavefunction! 
This corresponds, in the associated classical system, to being able to change 
the stochastic dynamics without affecting the equilibrium partition function. 
We will also establish that representations of quantum Hamiltonians, 
in some preferred basis 
$\mathcal{B}=\{|\mathcal{C}\rangle, \; \mathcal{C}\in\mathcal{S}\}$, 
that are SMF decomposable 
are in one-to-one correspondence with discrete classical systems 
with configuration space $\mathcal{S}$ endowed with 
stochastic dynamics described by a Master equation of the matrix type. 
The dictionary between the quantum and classical systems is 
summarized in Table~\ref{table: Dictionary between quantum and classical}. 
\begin{table}[t]
\caption{
\label{table: Dictionary between quantum and classical}
Correspondences between the SMF representation of the quantum 
Hamiltonian $\widehat{H}^{\ }_{SMF}$ 
from Eq.\ (\ref{eq: H RK positive def})
and the classical system with stochastic 
dynamics in time captured by the transition matrix 
$(W^{\ }_{\mathcal{C}\mathcal{C}'})$ 
from Eq.\ (\ref{eq: def tilde W and W}). 
\vspace{0.4 cm}}
\begin{center}
\begin{tabular*}{\textwidth}{@{\extracolsep{\fill}}llll}
\hline\hline
\hspace{0.03\textwidth}
&
Quantum system admitting an SMF decomposition
&
Classical system
&
\hspace{0.00\textwidth}
\\\hline\vspace{-0.3 cm}\\
\hspace{0.03\textwidth}
&
Hilbert space with basis $\mathcal{B}$ labeled by $\mathcal{S}$
\hspace{40 pt} &
Configuration space $\mathcal{S}$
&
\hspace{0.00\textwidth}
\\
\hspace{0.03\textwidth}
&
Ground state wavefunction
&
Boltzmann distribution
&
\hspace{0.00\textwidth}
\\
\hspace{0.03\textwidth}
&
Quantum phase transitions
&
Classical phase transitions
&
\hspace{0.00\textwidth}
\\
\hspace{0.03\textwidth}
&
Hamiltonian matrix: 
$\left(\langle\mathcal{C}| \widehat{H}^{\ }_{SMF} |\mathcal{C}'\rangle\right)$
&
Transition matrix: 
$\left(W^{\ }_{\mathcal{C}\mathcal{C}'}\right)$
&
\hspace{0.00\textwidth}
\vspace{0.2 cm}\\
\hspace{0.03\textwidth}
&
\begin{math}
\left.
  \begin{array}{l}
    \textrm{\!\!Positive-semidefinite}
    \vspace{-0.28 cm}\\
    \textrm{\;\;\;\; decomposition conditions}
    \vspace{-0.0 cm}\\
    \textrm{\!\!Integrability conditions}
  \end{array}
\right\}
\end{math}
\hspace{10 pt}
&
\begin{math}
\left\{
  \begin{array}{l}
    \textrm{\!Positive transition rates}
    \vspace{0.38 cm}\\
    \textrm{\!Conservation of probabilities}
  \end{array}
\right.
\end{math}
\hspace{10 pt}
\vspace{0.2 cm}\\
\hspace{0.00\textwidth}
&
Energy eigenvalues
&
Relaxation rates
&
\hspace{0.00\textwidth}
\\
\hspace{0.03\textwidth}
&
Eigenfunctions
&
Right/Left eigenfunctions
&
\hspace{0.00\textwidth}
\\\hline\hline
\end{tabular*}
\end{center}
\end{table}

So far, our generalization of the known quantum Hamiltonians 
that are fine-tuned to their RK points is 
predicated on choosing a preferred basis first, i.e., given a special choice 
of basis, we have investigated under what conditions a quantum 
Hamiltonian is SMF decomposable. This is so for historical reasons. 
One can of course ask the reverse question, namely 
is there a basis in which some given quantum Hamiltonian 
is SMF decomposable? 
We shall give an affirmative answer to this question for any 
quantum Hamiltonian defined on a finite dimensional Hilbert space 
as long as it has not too many degenerate eigenvalues, 
up to some trivial shift of the energy spectrum. 
In fact, there are continuously many bases in which such a quantum 
Hamiltonian admits distinct representations that are SMF decomposable. 
The correspondence between a quantum Hamiltonian, when understood as 
an abstract operator acting on some Hilbert space, 
and stochastic classical systems is thus one-to-(continuously) many. 

As an application of these results, we shall illustrate a systematic 
procedure to include dimer-dimer, vertex-vertex, or color-color interactions 
in known quantum models that are tuned to their RK points, 
such as the quantum dimer, vertex, and color models, respectively. 
In particular, we discuss the procedure in greater detail for the case of the 
quantum three-coloring model. These extra interactions can lead to additional 
quantum phases and, for quantum Hamiltonians that are SMF decomposable, 
such quantum phases and quantum phase transitions can be partly understood 
in terms of purely classical phase transitions at thermal equilibrium. 
A second facet of these results is that the quantum Hamiltonians have simple 
representations as sums over ``elementary'' operators with a rather intuitive 
interpretation in terms of ``elementary'' moves in the associated classical 
configuration spaces. For example, in the quantum three-coloring model, we 
shall first identify ``decorated'' loops in the classical version of the 
model, from which we construct the quantum Hamiltonian as a sum over 
``elementary'' operators associated to these ``decorated'' objects. Each of 
these ``elementary'' operators corresponds to an ``elementary'' move in the 
associated classical phase space that is energy non-conserving. A similar 
procedure can be carried out in quantum dimer and vertex models after having 
identified ``decorated'' plaquettes that encode non-trivial dimer-dimer and 
vertex-vertex interactions, respectively. 

The first main result of our paper, in the form of the generic conditions 
on a given representation of a quantum Hamiltonian for it to be SMF 
decomposable, is derived in Sec.~\ref{sec: Generalized quantum dimer models}. 
In Sec.~\ref{sec: From quantum to classical dynamics} we show how the 
quantum dynamics of a quantum Hamiltonian that admits a representation
that is SMF decomposable induce classical stochastic 
dynamics encoded by an appropriate Master equation. 
This result, when combined with 
Henley's result in the opposite direction,~\cite{Henley04} 
allows us to establish the one-to-one correspondence between 
all possible SMF decompositions of quantum Hamiltonians 
in the preferred basis 
$\mathcal{B}=\{|\mathcal{C}\rangle, \; \mathcal{C}\in\mathcal{S}\}$ 
and discrete, stochastic classical systems described by a Master equation 
of the matrix type 
in the classical configuration space $\mathcal{S}$. 
In Sec.~\ref{sec: Beyond RK Hamiltonians: The Stochastic Matrix Form} we 
present the second main result of the paper, namely the conditions under which
a quantum Hamiltonian admits a continuous manifold of bases for which
it is SMF decomposable. 
Specific examples of how known quantum Hamiltonians that are fine-tuned to
their RK points can be understood 
within our framework and how this allows for immediate generalizations are 
given in Sec.~\ref{sec: Examples}. These examples are the triangular dimer 
model, the eight-vertex model, and the three-coloring model on the honeycomb 
lattice. Conclusions are drawn in Sec.~\ref{sec: Conclusions}. 

\section{The Stochastic Matrix Form decomposition of quantum Hamiltonians}
\label{sec: Generalized quantum dimer models} 

In this section we are going to construct the most general form of a 
quantum Hamiltonian, when represented in a specific basis, satisfying 
the properties~(I-III). 

Consider the Hilbert space $\mathcal{H}$ given by the span of all orthogonal 
and normalized (orthonormal) states 
$|\mathcal{C}\rangle$, labeled by the index 
$\mathcal{C}\in\mathcal{S}$ with $\mathcal{S}$ a 
countable set (condition (I)), 
\begin{equation}
\label{eq: def basis Hilbert space}
  \langle\mathcal{C}|\mathcal{C}'\rangle = 
    \delta^{\ }_{\mathcal{C}\mathcal{C}'}, 
  \qquad 
  \openone 
  = \sum_{\mathcal{C}\in\mathcal{S}} 
    |\mathcal{C}\rangle\langle\mathcal{C}|.
\end{equation}
Consider a generic (Hermitian) Hamiltonian $\widehat{H}$ acting on this 
Hilbert space, and define the subset 
$\mathcal{L}\subset\mathcal{S}\times\mathcal{S}$ to be the set of all the 
pairs $(\mathcal{C},\mathcal{C}')$ 
with $\mathcal{C} \neq \mathcal{C}'$ 
such that the off-diagonal matrix elements of the 
$|\mathcal{S}|\times|\mathcal{S}|$ Hermitian matrix 
\begin{equation} 
\left(H^{\ }_{\mathcal{C}\mathcal{C}'}\right):=
\left(\langle\mathcal{C}|\widehat{H}|\mathcal{C}'\rangle\right)
\end{equation} 
are non-vanishing. 
For simplicity, we will make the two technical assumptions that the Hilbert 
space is finite dimensional and fully connected under the time-evolution 
operator, i.e., any two states in $\mathcal{H}$ have a non-vanishing matrix 
element for some power of $\widehat{H}$. 
These two conditions will be needed when using the Perron-Frobenius 
theorem to establish the non-degeneracy of the GS. 
Then, we can always represent $\widehat{H}$ in the basis 
$\mathcal{B}=\{|\mathcal{C}\rangle, \; \mathcal{C}\in\mathcal{S}\}$ by%
~\cite{note: LS} 
\begin{subequations}
\label{eq: def H}
  \begin{equation}
  \label{eq: def H a}
    \widehat{H} = 
    \frac{1}{2}
    \sum_{(\mathcal{C},\mathcal{C}')\in \mathcal{L}}
    \widehat{h}^{\ }_{\mathcal{C},\mathcal{C}'}
  \end{equation}
where 
  \begin{equation}
  \label{eq: def H b}
\begin{split}
    \widehat{h}^{\ }_{\mathcal{C},\mathcal{C}'} &:=
    \alpha^{\ }_{\mathcal{C},\mathcal{C}'}
    |\mathcal{C}\rangle\langle\mathcal{C}|
    +
    \alpha^{\ }_{\mathcal{C}',\mathcal{C}}
    |\mathcal{C}'\rangle\langle\mathcal{C}'|
    -
    \gamma^{\ }_{\mathcal{C},\mathcal{C}'}
    |\mathcal{C}\rangle\langle\mathcal{C}'|
    -
    \gamma^{\ }_{\mathcal{C}',\mathcal{C}}
    |\mathcal{C}'\rangle\langle\mathcal{C}|
\end{split}
  \end{equation}
and 
\begin{equation}
\alpha^{\ }_{\mathcal{C},\mathcal{C}'}\in\mathbb{R},\qquad 
\gamma^{\ }_{\mathcal{C},\mathcal{C}'}=
\gamma^{* }_{\mathcal{C}',\mathcal{C}}\in\mathbb{C}
\end{equation}
\end{subequations}
($\gamma^{* }_{\mathcal{C},\mathcal{C}'}$ is the complex conjugate to 
$\gamma^{\ }_{\mathcal{C},\mathcal{C}'}$). 
Observe that the representation (\ref{eq: def H}) is redundant. Indeed, 
the matrix elements of $\widehat{H}$ in this preferred basis, 
\begin{equation}
\label{eq: H elements a}
H^{\ }_{\mathcal{C}\mathcal{C}} = 
\sum_{\mathcal{C}'\in\mathcal{S}}^{(\mathcal{C},\mathcal{C}')\in\mathcal{L}}
\alpha^{\ }_{\mathcal{C},\mathcal{C}'},
\qquad\forall\,\mathcal{C}\in\mathcal{S},
\end{equation}
\begin{equation}
\label{eq: H elements b}
H^{\ }_{\mathcal{C}\mathcal{C}'} = 
-
\gamma^{\ }_{\mathcal{C} ,\mathcal{C}'},
\qquad\forall\,(\mathcal{C},\mathcal{C}')\in\mathcal{L},
\end{equation}
are unchanged under the transformations 
\begin{equation}
\begin{split}
&
\alpha^{\ }_{\mathcal{C} ,\mathcal{C}'} 
\;\rightarrow \;
\widetilde{\alpha}^{\ }_{\mathcal{C} ,\mathcal{C}'} = 
\alpha^{\ }_{\mathcal{C} ,\mathcal{C}'} 
+
f^{\ }_{\mathcal{C} ,\mathcal{C}'},
\end{split}
\end{equation}
provided 
\begin{equation}
\label{eq: alpha transf cond}
0=
\sum_{\mathcal{C}'\in\mathcal{S}}^{(\mathcal{C},\mathcal{C}')\in\mathcal{L}}
f^{\ }_{\mathcal{C} ,\mathcal{C}'},
\qquad\forall\,\mathcal{C}\in\mathcal{S}, 
\end{equation}
where the two summations, in Eqs.~(\ref{eq: H elements a}) 
and~(\ref{eq: alpha transf cond}), are over the single index 
$\mathcal{C}'\in\mathcal{S}$ subject to the constraint that
$(\mathcal{C},\mathcal{C}')\in\mathcal{L}$.
This is so because our Hermitian matrix 
$\left(H^{\ }_{\mathcal{C}\mathcal{C}'}\right)$ has 
$|\mathcal{S}|$ independent real-valued diagonal entries and 
$|\mathcal{L}|/2$ non-vanishing independent off-diagonal complex-valued 
matrix elements. On the other hand, there are 
$|\mathcal{L}|$ independent 
$\alpha$'s and $|\mathcal{L}|/2$ independent $\gamma$'s. 
The representation~(\ref{eq: def H}) is thus redundant, at least for 
$|\mathcal{S}|>2$, since $|\mathcal{L}|\geq2(|\mathcal{S}|-1)$ 
(see~\cite{note: LS}). 
It will nevertheless be useful to identify the set of data 
$(\mathcal{L},
  \alpha^{\ }_{\mathcal{C},\mathcal{C}'},
  \gamma^{\ }_{\mathcal{C},\mathcal{C}'})
$ 
with the Hamiltonian $\widehat{H}$. 

Notice the competition between the potential energy, the first two terms of 
Eq.~(\ref{eq: def H b}), and the kinetic energy, the last two terms of 
Eq.~(\ref{eq: def H b}), in each Hermitian operator 
$\widehat{h}^{\ }_{\mathcal{C},\mathcal{C}'}$. 
In order to satisfy condition (II), we have to require that 
\begin{equation}
\label{eq: condition for positive definite}
  \alpha^{\ }_{\mathcal{C},\mathcal{C}'}\times 
    \alpha^{\ }_{\mathcal{C}',\mathcal{C}} = 
      |\gamma^{\ }_{\mathcal{C},\mathcal{C}'}|^2 
  \;\;\; \textrm{and} \;\;\;
  \alpha^{\ }_{\mathcal{C},\mathcal{C}'}
    > 0 
\end{equation}
for all $(\mathcal{C},\mathcal{C}')\in\mathcal{L}$, 
in which case we are left with $|\mathcal{L}|/2\geq(|\mathcal{S}|-1)$ 
independent $\alpha$'s. 
Indeed, one verifies that Eqs.~(\ref{eq: def H b}) 
and~(\ref{eq: condition for positive definite}) imply 
\begin{equation}
\label{eq: h square propto h}
  \widehat{h}^{2 }_{\mathcal{C},\mathcal{C}'} = 
    \left( 
      \alpha^{\ }_{\mathcal{C},\mathcal{C}'} + 
      \alpha^{\ }_{\mathcal{C}',\mathcal{C}} 
    \right)
  \widehat{h}^{\ }_{\mathcal{C},\mathcal{C}'}. 
\end{equation}
The first condition in Eq.~(\ref{eq: condition for positive definite}) 
amounts to nothing else but imposing that the $2\times2$ Hermitian matrix 
\begin{equation}
\left( 
\begin{array}{cc} 
\alpha^{\ }_{\mathcal{C},\mathcal{C}'}&
\gamma^{\ }_{\mathcal{C},\mathcal{C}'}\\
\gamma^{* }_{\mathcal{C},\mathcal{C}'}&
\alpha^{\ }_{\mathcal{C}',\mathcal{C}}
\end{array}
\right), 
\end{equation} 
defined on the two-dimensional vector space spanned by $|\mathcal{C}\rangle$ 
and $|\mathcal{C}'\rangle$, has vanishing determinant. This is a necessary 
and sufficient condition for a Hermitian $2\times2$ matrix 
to be proportional to a projection operator. 
Together, the two equations in~(\ref{eq: condition for positive definite}) 
form the positive-semidefinite decomposition conditions, that are necessary 
and sufficient for the Hamiltonian $\widehat{H}$ to be decomposable in 
$2\times2$ Hermitian blocks, 
each of which has precisely one null and one positive eigenvalue. 
Equation~(\ref{eq: condition for positive definite}) is also sufficient to 
guarantee that $\widehat{H}$ is a positive semidefinite quantum Hamiltonian. 
Any eigenstate annihilated by $\widehat{H}$ thus necessarily belongs to the 
GS manifold whenever 
conditions~(\ref{eq: condition for positive definite}) hold. 

Before imposing condition~(III), it is useful to change the parametrization 
in Eq.~(\ref{eq: def H b}) to 
\begin{subequations}
\label{eq: new parameters}
  \begin{equation}
    w^{\ }_{\mathcal{C},\mathcal{C}'} := 
      \sqrt{\alpha^{\ }_{\mathcal{C},\mathcal{C}'}\times
            \alpha^{\ }_{\mathcal{C}',\mathcal{C}}}, 
  \end{equation}
  \begin{equation}
    e^{-K\varepsilon^{\ }_{\mathcal{C},\mathcal{C}'}/2} := 
      \sqrt{\frac{\alpha^{\ }_{\mathcal{C},\mathcal{C}'}}
                 {\alpha^{\ }_{\mathcal{C}',\mathcal{C}}}}, 
  \end{equation}
  \begin{equation}
     K\phi^{\ }_{\mathcal{C},\mathcal{C}'}/2 :=  
      \arg(\gamma^{\ }_{\mathcal{C},\mathcal{C}'}). 
  \end{equation}      
\end{subequations}
Here, the coupling constants $K\varepsilon^{\ }_{\mathcal{C},\mathcal{C}'}$ 
and $K\phi^{\ }_{\mathcal{C},\mathcal{C}'}$ are real valued and the 
$w^{\ }_{\mathcal{C},\mathcal{C}'}$ are strictly positive. 
It is useful, even though redundant, to represent the real coupling constants 
$K\varepsilon^{\ }_{\mathcal{C},\mathcal{C}'}$ and 
$K\phi^{\ }_{\mathcal{C},\mathcal{C}'}$ as products of a common scaling 
factor $K$ and two state-dependent functions 
$\varepsilon^{\ }_{\mathcal{C},\mathcal{C}'}$ and 
$\phi^{\ }_{\mathcal{C},\mathcal{C}'}$. 
The reasons for this choice will become clear shortly, and rely on the 
physical interpretation of the scaling factor $K$ (see 
Eq.~(\ref{eq: physical interp of K})). 
After this change in parametrization, Eq.~(\ref{eq: def H}) becomes 
\begin{subequations}
\label{eq: H positive def}
  \begin{equation}
  \label{eq: H positive def a}
    \widehat{H} = 
      \frac{1}{2}
      \sum_{(\mathcal{C},\mathcal{C}')\in\mathcal{L}} 
      w^{\ }_{\mathcal{C},\mathcal{C}'}\, 
        \widehat{Q}^{\ }_{\mathcal{C},\mathcal{C}'}
  \end{equation}
where 
  \begin{equation}
  \label{eq: H positive def b}
\begin{split}
    \widehat{Q}^{\ }_{\mathcal{C},\mathcal{C}'} &= 
      e^{-K\varepsilon^{\ }_{\mathcal{C},\mathcal{C}'}/2} 
        |\mathcal{C}\rangle\langle\mathcal{C}|
      +e^{+K\varepsilon^{\ }_{\mathcal{C},\mathcal{C}'}/2}
        |\mathcal{C}'\rangle\langle\mathcal{C}'|
      \\
      &- 
      e^{-iK\phi^{\ }_{\mathcal{C},\mathcal{C}'}/2}\, 
        |\mathcal{C}'\rangle\langle\mathcal{C}|
      -e^{+iK\phi^{\ }_{\mathcal{C},\mathcal{C}'}/2}\, 
        |\mathcal{C}\rangle\langle\mathcal{C}'|. 
\end{split}
  \end{equation}
\end{subequations}
We now require that condition (III) be satisfied, i.e., that there 
exists a simultaneous zero mode for all the Hermitian operators 
$\widehat{Q}^{\ }_{\mathcal{C},\mathcal{C}'}$ in 
Eq.~(\ref{eq: H positive def b}). 
One verifies that the nodeless wavefunction 
\begin{equation}
\label{zero-mode} 
  |\Psi^{\ }_{0}\rangle := \sum_{\mathcal{C}\in\mathcal{S}} 
    e^{-KE^{\ }_{\mathcal{C}}/2} |\mathcal{C}\rangle, 
\end{equation}
is annihilated by $\widehat{Q}^{\ }_{\mathcal{C},\mathcal{C}'}$, 
$\forall\,(\mathcal{C},\mathcal{C}')\in\mathcal{L}$, 
provided the integrability conditions 
\begin{subequations}
\label{eq: integrability conds}
  \begin{equation}
  \label{eq: integrability conds a}
    \varepsilon^{\ }_{\mathcal{C},\mathcal{C}'} = 
      E^{\ }_{\mathcal{C}'} - E^{\ }_{\mathcal{C}} 
  \end{equation}
and 
  \begin{equation}
  \label{eq: integrability conds b}
    \phi^{\ }_{\mathcal{C},\mathcal{C}'} = 0 
    \;\;\;\; \hbox{\textit{(up to a pure gauge transformation)}}
  \end{equation}
\end{subequations}

\noindent
on the real-valued parameters $\varepsilon^{\ }_{\mathcal{C},\mathcal{C}'}$ 
and $\phi^{\ }_{\mathcal{C},\mathcal{C}'}$ are satisfied for some 
real-valued function $E^{\ }_{\mathcal{C}}$ defined on $\mathcal{S}$, 
and for all $(\mathcal{C},\mathcal{C}')\in\mathcal{L}$. 
As a result of the integrability conditions~(\ref{eq: integrability conds b}) 
and noting that all the couplings $w^{\ }_{\mathcal{C},\mathcal{C}'}$ 
are positive by definition, we see that all off-diagonal matrix elements of 
$\widehat{H}$ are negative. This property, together with the integrability 
conditions~(\ref{eq: integrability conds a}) and the positive-semidefinite 
decomposition conditions~(\ref{eq: condition for positive definite}), 
will be needed to infer that $\widehat{H}$ admits an SMF decomposition. 
Here, it suffices to say that the two 
conditions~(\ref{eq: integrability conds}) 
guarantee that $|\Psi^{\ }_{0}\rangle$ is the GS of the Hamiltonian 
$\widehat{H}$ in Eq.~(\ref{eq: H positive def a}), 
regardless of the values taken 
by the coupling constants $w^{\ }_{\mathcal{C},\mathcal{C}'}$, 
and that this GS is non-degenerate by the Perron-Frobenius 
theorem~\cite{Perron-Frobenius}. 

We conclude that the most general form for a quantum Hamiltonian 
that is defined on the Hilbert space spanned by the basis 
$\mathcal{B}=\{|\mathcal{C}\rangle, \; \mathcal{C}\in\mathcal{S}\}$ 
and that satisfies the RK conditions~(I-III) 
is to depend on the data  
$(\mathcal{L},
  w^{\ }_{\mathcal{C},\mathcal{C}'},
  E^{\ }_{\mathcal{C}},
  K)
$ 
through the representation 
\begin{subequations}
\label{eq: H RK positive def}
  \begin{equation}
  \label{eq: H RK positive def a}
    \widehat{H}^{\ }_{SMF} = 
      \frac{1}{2}
      \sum_{(\mathcal{C},\mathcal{C}')\in\mathcal{L}} 
      w^{\ }_{\mathcal{C},\mathcal{C}'}\, 
      \widehat{Q}^{\ }_{\mathcal{C},\mathcal{C}'}
  \end{equation}
where~\cite{note: 2x2 RK matrix} 
  \begin{equation}
  \label{eq: H RK positive def b}
\begin{split}
    \widehat{Q}^{\ }_{\mathcal{C},\mathcal{C}'} &=
      e^{-K (E^{\ }_{\mathcal{C}'}-E^{\ }_{\mathcal{C}})/2} 
        |\mathcal{C}\rangle\langle\mathcal{C}| 
      +
      e^{+K (E^{\ }_{\mathcal{C}'}-E^{\ }_{\mathcal{C}})/2}
        |\mathcal{C}'\rangle\langle\mathcal{C}'|
    -
    |\mathcal{C}'\rangle\langle\mathcal{C}|
    -
    |\mathcal{C}\rangle\langle\mathcal{C}'|. 
\end{split}
  \end{equation}
This representation will be referred to hereafter as the 
\textit{Stochastic Matrix Form} (SMF) decomposition.
The normalized GS wavefunction 
  \begin{equation}
  \label{eq: H RK ground state c}
    |\Psi^{\ }_{SMF}\rangle = \frac{1}{\sqrt{Z(K)}} 
      \sum_{\mathcal{C}\in\mathcal{S}} e^{-KE^{\ }_{\mathcal{C}}/2} 
        |\mathcal{C}\rangle 
  \end{equation}
of $\widehat{H}^{\ }_{SMF}$ 
is non-degenerate as long as the Hilbert space is 
connected under the time-evolution operator and finite-dimensional.
Here, the normalization factor 
  \begin{equation}
  \label{eq: def Z(K)}
    Z(K):= \sum_{\mathcal{C}\in\mathcal{S}} 
      e^{-K E^{\ }_{\mathcal{C}}} 
  \end{equation}
can be interpreted as the partition function of a classical system with 
phase space $\mathcal{S}$, reduced inverse temperature 
  \begin{equation}
  \label{eq: physical interp of K}
    K:=\beta J, 
  \end{equation}
\end{subequations}
($J$ being some characteristic energy scale), and dimensionless energies 
$E^{\ }_{\mathcal{C}}$. Evidently, in the infinite temperature limit 
($\beta=0$) and in a finite system, $Z(0)$ reduces to the combinatorial 
problem of counting the number of classical configurations belonging to 
$\mathcal{S}$. 

It is instructive to observe that Eq.~(\ref{eq: H elements a}) 
now takes the form 
\begin{subequations}
\label{eq: diagonal in terms off diagonal}
\begin{equation}
\label{eq: diagonal in terms off diagonal a}
\begin{split}
H^{\ }_{\mathcal{C}\mathcal{C}}&= 
\sum_{\mathcal{C}'\in\mathcal{S}}^{(\mathcal{C},\mathcal{C}')\in\mathcal{L}}
w^{\ }_{\mathcal{C},\mathcal{C}'}\, 
e^{-K\varepsilon^{\ }_{\mathcal{C},\mathcal{C}'}/2}
\\
&=
\sum_{\mathcal{C}'\in\mathcal{S}}^{(\mathcal{C},\mathcal{C}')\in\mathcal{L}}
|H^{\ }_{\mathcal{C}\mathcal{C}'}|\,
e^{-K\varepsilon^{\ }_{\mathcal{C},\mathcal{C}'}/2},
\\
&=
\sum_{\mathcal{C}'\in\mathcal{S}}^{(\mathcal{C},\mathcal{C}')\in\mathcal{L}}
|H^{\ }_{\mathcal{C}\mathcal{C}'}|\,
\frac{\Psi^{\ }_{0,\mathcal{C}}}{\Psi^{\ }_{0,\mathcal{C}'}}, 
\qquad\forall\,\mathcal{C}\in\mathcal{S},
\end{split}
\end{equation}
where we have introduced the notation 
\begin{equation}
\label{eq: diagonal in terms off diagonal b}
\Psi^{\ }_{0,\mathcal{C}}:=
  \langle\mathcal{C}|\Psi^{\ }_{0}\rangle = 
    \frac{e^{-KE^{\ }_{\mathcal{C}}/2}}{\sqrt{Z(K)}}. 
\end{equation} 
\end{subequations}
As we shall prove in 
Sec.~\ref{sec: Beyond RK Hamiltonians: The Stochastic Matrix Form}, 
for any quantum Hamiltonian $\widehat{H}$ that has a ground state 
$|\Psi^{\ }_{0}\rangle$ with vanishing eigenvalue, and that admits an 
irreducible representation $(H^{\ }_{\mathcal{C}\mathcal{C}'})$ with 
non-positive off-diagonal matrix elements in some finite dimensional basis 
$\mathcal{B}=\{|\mathcal{C}\rangle, \; \mathcal{C}\in\mathcal{S}\}$, 
it is possible to define in a unique way the 
decomposition~(\ref{eq: H positive def}) with all 
$\phi^{\ }_{\mathcal{C},\mathcal{C}'}=0$, by defining the sets 
$\{K\varepsilon^{\ }_{\mathcal{C},\mathcal{C}'},\;
   (\mathcal{C},\mathcal{C}')\in\mathcal{L}\}$ 
and 
$\{w^{\ }_{\mathcal{C},\mathcal{C}'},\;
   (\mathcal{C},\mathcal{C}')\in\mathcal{L}\}$ 
through $K\varepsilon^{\ }_{\mathcal{C},\mathcal{C}'}/2 := 
\ln(\Psi^{\ }_{0,\mathcal{C}'}/\Psi^{\ }_{0,\mathcal{C}})$ and through 
$w^{\ }_{\mathcal{C},\mathcal{C}'} := |H^{\ }_{\mathcal{C}\mathcal{C}'}|$, 
so that Eq.~(\ref{eq: diagonal in terms off diagonal a}) is satisfied. 
We are now equipped with \textit{necessary and sufficient conditions} 
for the matrix representation of a quantum Hamiltonian 
to be uniquely decomposable in the 
SMF~(\ref{eq: H RK positive def}). 

Equations~(\ref{eq: H RK positive def}) encode our first important result: 
\textit{
Given any quantum problem and a preferred basis, 
i.e., given a Hilbert space $\mathcal{H}$ and a set of data 
$(\mathcal{L},
  \alpha^{\ }_{\mathcal{C},\mathcal{C}'},
  \gamma^{\ }_{\mathcal{C},\mathcal{C}'})
$ 
as defined above, it satisfies the RK conditions (I-III) 
\textbf{iff} 
$\mathcal{H}$ is separable and its coupling constants 
$\alpha^{\ }_{\mathcal{C},\mathcal{C}'}$ and 
$\gamma^{\ }_{\mathcal{C},\mathcal{C}'}$ satisfy 
the positive-semidefinite decomposition 
conditions~(\ref{eq: condition for positive definite}) 
and the integrability conditions~(\ref{eq: integrability conds}).}
The coupling constants can then be redefined 
\begin{equation}
(\mathcal{L},
  \alpha^{\ }_{\mathcal{C},\mathcal{C}'},
  \gamma^{\ }_{\mathcal{C},\mathcal{C}'}
)
\to
(\mathcal{L},
  w^{\ }_{\mathcal{C},\mathcal{C}'},
  E^{\ }_{\mathcal{C}},
  K
)
\end{equation}
and the Hamiltonian representation takes the simpler 
form~(\ref{eq: H RK positive def}). 
One can associate this quantum problem to a classical problem in
thermal equilibrium described by 
the partition function~(\ref{eq: def Z(K)}) 
appearing in the GS wavefunction~(\ref{eq: H RK ground state c}) of 
$\widehat{H}^{\ }_{SMF}$. 
All equal-time quantum correlation functions for operators diagonal in the 
preferred basis~(\ref{eq: def basis Hilbert space}) of $\mathcal{H}$ 
are identical to their classical counterparts in thermal equilibrium at 
reduced inverse temperature $K$. Any finite temperature phase transition, say 
at $K^{\ }_c$, of the classical model~(\ref{eq: def Z(K)}) implies a quantum 
(zero-temperature) phase transition at the RK point parametrized by the 
quantum coupling constant $K^{\ }_c$ (the converse statement is not true!). 
Another remarkable property of $\widehat{H}^{\ }_{SMF}$ is that, while the 
GS wavefunction is completely independent of the choice of the parameters 
$w^{\ }_{\mathcal{C},\mathcal{C}'}$, they instead control 
to a large extent the nature of the excitations above the 
GS~(\ref{eq: H RK ground state c}). 

Conversely, for any discrete classical statistical system in thermal 
equilibrium as defined by the partition function~(\ref{eq: def Z(K)}), 
say, we can construct the Hamiltonian $\widehat{H}^{\ }_{SMF}$ 
with GS wavefunction~(\ref{eq: H RK ground state c}). 
Of course, $\widehat{H}^{\ }_{SMF}$ is not uniquely defined as the coupling 
constants $w^{\ }_{\mathcal{C},\mathcal{C}'}$ in the set of data 
$(\mathcal{L},
  w^{\ }_{\mathcal{C},\mathcal{C}'},
  E^{\ }_{\mathcal{C}},
  K)
$ 
needed to specify uniquely $\widehat{H}^{\ }_{SMF}$ 
cannot be extracted from the equilibrium properties of the classical system. 
For the full correspondence between quantum Hamiltonian representations that 
are SMF decomposable and classical statistical systems to be established, 
we need also to account for the approach to thermal equilibrium in the 
classical system as we now explain. 

\section{From quantum dynamics to classical stochastics}
\label{sec: From quantum to classical dynamics}

Henley observed that the eigenvalues of the square lattice quantum 
dimer Hamiltonian at its RK point are identical to the relaxation rates of 
a (classical) Master equation~\cite{Henley97}. 
Ivanov took advantage of this observation to simulate the vison gap of the 
triangular lattice quantum dimer Hamiltonian at its RK point through a 
classical Monte Carlo simulation on the same lattice 
as the one for the quantum model~\cite{Ivanov04}. 
Henley also observed that any generic Master equation can be brought 
to the form of a quantum Hamiltonian fine-tuned to its RK point 
through a similarity transformation in a unique fashion~\cite{Henley04}. 
The reverse question of whether any quantum Hamiltonian, when fine-tuned
to an RK point, can be brought to the form of a Master equation has 
not yet been addressed. 

In this section we will construct a (classical) Master equation from any 
given quantum Hamiltonian that admits an SMF decomposition
when represented in a preferred basis, 
such that its stationary probability distribution is nothing but the 
Boltzmann weights from Eq.~(\ref{eq: def Z(K)}). 
This Master equation can be thought of as the result of an environment that 
endows the classical system with specific stochastics in time. 
We will then show that the positive-semidefinite decomposition 
conditions~(\ref{eq: condition for positive definite}) 
and the integrability conditions~(\ref{eq: integrability conds}) 
combine in such a way that the Master equation can be written in matrix form, 
and the eigenvalues of the transition matrix (i.e., the relaxation 
rates of the classical system) are identical, up to an overall sign, to the 
positive eigenvalues of the quantum Hamiltonian~(\ref{eq: H RK positive def}). 
Here, the (local) positive-semidefinite decomposition 
conditions~(\ref{eq: condition for positive definite}) 
are sufficient for all eigenvalues of the transition matrix to be 
non-positive, and together with 
the (global) integrability conditions~(\ref{eq: integrability conds}) 
they are necessary for the transition rates to be non-negative and for the 
eigenstates of the transition matrix 
to have a conserved normalization under time evolution (i.e., 
conservation of probability holds). 
This establishes the correspondence between 
quantum dynamics and classical stochastics 
in the complementary direction to the one explored by Henley. 
The general character of the classical systems associated to SMF 
representations of quantum Hamiltonians allows us to present the 
following result: 
\textit{%
There exists a one-to-one correspondence between quantum Hamiltonians 
that admit an SMF decomposition~(\ref{eq: H RK positive def}) 
in a given preferred basis 
$\mathcal{B}=\{|\mathcal{C}\rangle, \; \mathcal{C}\in\mathcal{S}\}$
and classical statistical systems endowed with time stochastics through a 
Master equation of the matrix 
type~(\ref{eq: Master equations as a matrix equation})
in the given configuration space $\mathcal{S}$.} 

\subsection{Master equation and transition matrix}

From now on, we denote the eigenvalues and the orthonormal
eigenstates of $\widehat{H}^{\ }_{SMF}$ in Eq.~(\ref{eq: H RK positive def}) 
by $\varepsilon^{\ }_{n}$ and $|\varepsilon^{\ }_{n}\rangle$ respectively, 
i.e., 
\begin{equation}
\label{eq: notation for spectral decomposition HRK}
  \widehat{H}^{\ }_{SMF} = \sum_{n} \varepsilon^{\ }_{n}\, 
    |\varepsilon^{\ }_{n}\rangle \langle\varepsilon^{\ }_{n}|
\end{equation}
with the non-degenerate GS $|\varepsilon^{\ }_{n=0}\rangle \equiv 
|\Psi^{\ }_{SMF}\rangle$. 
Choose any two $\mathcal{C}$ and $\mathcal{C}'$ in $\mathcal{S}$ 
and define the matrix elements 
\begin{subequations}
\label{eq: def tilde W and W}
  \begin{equation}
  \label{eq: def tilde W and W a}
    H^{\ }_{\mathcal{C}\mathcal{C}'} := 
    \langle\mathcal{C}|
    \widehat H^{\ }_{SMF}
    |\mathcal{C}'\rangle
  \end{equation}
and 
  \begin{equation}
  \label{eq: def tilde W and W b}
    W^{\ }_{\mathcal{C}\mathcal{C}'} := 
    -
    e^{-
    K\left(E^{\ }_{\mathcal{C}}-E^{\ }_{\mathcal{C}'}\right)/2
      }
    H^{\ }_{\mathcal{C}\mathcal{C}'},
  \end{equation}
\end{subequations}
respectively. Hermiticity and time-reversal symmetry of 
$\widehat{H}^{\ }_{SMF}$ imply the condition of symmetry 
\begin{subequations}
\label{eq: symmetry tilde W and detailed balance W} 
  \begin{equation}
  \label{eq: symmetry tilde W and detailed balance W a} 
    H^{\ }_{\mathcal{C}\mathcal{C}'} = 
    H^{\ }_{\mathcal{C}'\mathcal{C}}
  \end{equation}
which, in turn, implies the condition of detailed balance 
  \begin{equation} 
  \label{eq: symmetry tilde W and detailed balance W b} 
    W^{\ }_{\mathcal{C}\mathcal{C}'} 
      e^{-K E^{\ }_{\mathcal{C}'}} = 
        W^{\ }_{\mathcal{C}'\mathcal{C}} 
          e^{-K E^{\ }_{\mathcal{C}}}
  \end{equation}
\end{subequations}
for any pair $\mathcal{C},\mathcal{C}'\in\mathcal{S}$. 

Because of the integrability conditions (\ref{eq: integrability conds b}), 
given Eqs.~(\ref{eq: new parameters},\ref{eq: H positive def}) and 
definitions~(\ref{eq: def tilde W and W}), 
all off-diagonal matrix elements in $(W^{\ }_{\mathcal{C}\mathcal{C}'})$ 
are positive and can thus be interpreted as transition rates (conditional 
probabilities). 
We are now in the position to define the Master equation 
  \begin{equation}
  \label{eq: Master equation a}
    \dot{p}^{\ }_{\mathcal{C}}(\tau) := 
      \sum_{\mathcal{C}'\in\mathcal{S}}
          ^{\mathcal{C}'\neq\mathcal{C}}
        \left[\vphantom{\Big[}
           W^{\ }_{\mathcal{C}\mathcal{C}'}
             p^{\ }_{\mathcal{C}'}(\tau) - 
               W^{\ }_{\mathcal{C}'\mathcal{C}}
	         p^{\ }_{\mathcal{C}}(\tau)
        \right],
     \qquad \forall\,\mathcal{C}\in\mathcal{S}
  \end{equation}
whose properly normalized solution can be interpreted as the instantaneous 
probability (a number between 0 and 1)
for the classical system to be in configuration $\mathcal{C}$. 
Equation~(\ref{eq: Master equation a}) defines in a natural and unique way 
the classical stochastics at the reduced temperature $K$ 
induced by the quantum Hamiltonian~(\ref{eq: H RK positive def}) 
on the associated classical system in thermal equilibrium. 
On the other hand, the special balance between the kinetic and the 
potential terms characteristic of a quantum Hamiltonian that admits
an SMF decomposition when represented in 
a preferred basis guarantees that the transition matrix 
$(W^{\ }_{\mathcal{C}\mathcal{C}'})$ satisfies the conservation of 
normalization condition 
%
%
$
  W^{\ }_{\mathcal{C}\mathcal{C}} = 
    -\sum_{\mathcal{C}'\in\mathcal{S}}
         ^{\mathcal{C}'\neq\mathcal{C}} 
       W^{\ }_{\mathcal{C}'\mathcal{C}}
$, 
$
\forall\,\mathcal{C}\in\mathcal{S}, 
$
%
%
as it can be verified directly using Eqs.~(\ref{eq: H RK positive def a},%
\ref{eq: H RK positive def b}) 
and~(\ref{eq: def tilde W and W b}). 
This is indeed a condition relating the diagonal and off-diagonal elements 
of $(W^{\ }_{\mathcal{C}\mathcal{C}'})$ 
which follows from the SMF  
conditions~(\ref{eq: condition for positive definite}) 
\textit{and}~(\ref{eq: integrability conds}). 
We are then allowed to recast Eq.~(\ref{eq: Master equation a}) 
as a Master equation of the matrix type 
\begin{subequations}
\label{eq: Master equations as a matrix equation}
  \begin{equation}
  \label{eq: Master equations as a matrix equation a}
    \dot{p}^{\ }_{\mathcal{C}}(\tau) = 
      \sum_{\mathcal{C}'\in\mathcal{S}}
        W^{\ }_{\mathcal{C}\mathcal{C}'}\,
          p^{\ }_{\mathcal{C}'}(\tau),
    \qquad\forall\,\mathcal{C}\in\mathcal{S},
  \end{equation}
where the (transition matrix) positivity conditions 
  \begin{equation}
  \label{eq: Master equations as a matrix equation b}
    W^{\ }_{\mathcal{C}\mathcal{C}'} \geq 0
  \qquad 
    \forall\,\mathcal{C},\mathcal{C}'\in\mathcal{S},\; 
      \mathcal{C}\neq\mathcal{C}',
  \end{equation}
the conservation of probability conditions 
  \begin{equation}
  \label{eq: Master equations as a matrix equation c}
    W^{\ }_{\mathcal{C}\mathcal{C}} = 
      -\sum_{\mathcal{C}'\in\mathcal{S}}
           ^{\mathcal{C}'\neq\mathcal{C}} 
         W^{\ }_{\mathcal{C}'\mathcal{C}}
    \qquad\forall\,\mathcal{C}\in\mathcal{S},
  \end{equation}
and the detailed balance conditions 
  \begin{equation} 
  \label{eq: Master equations as a matrix equation d}
    W^{\ }_{\mathcal{C}\mathcal{C}'}\, 
      p^{(0)}_{\mathcal{C}'}
    = 
    W^{\ }_{\mathcal{C}'\mathcal{C}}\, 
      p^{(0)}_{\mathcal{C}}
    \qquad\forall\,\mathcal{C},\mathcal{C}'\in\mathcal{S},
  \end{equation}
  \begin{equation} 
  \label{eq: Master equations as a matrix equation e}
    p^{(0)}_{\mathcal{C}} = \frac{1}{Z(K)}\, 
      e^{-K E^{\ }_{\mathcal{C}}}. 
    \qquad\qquad\qquad\;   
  \end{equation}
\end{subequations}
hold. 
The solutions to Eq.~(\ref{eq: Master equations as a matrix equation}) 
can be interpreted as probabilities at all times given that they are 
probabilities initially. They are of the form 
\begin{equation}
\label{eq: Solutions Master equations as a matrix equation}
  p^{\ }_{\mathcal{C}}(\tau) = \sum_{n\in\mathbb{N}} 
    a^{\ }_{n} e^{-|\lambda^{\ }_{n}|\tau} 
      \psi^{(R;n)}_{\mathcal{C}}, 
  \qquad
  a^{\ }_{n}\in\mathbb{R}
\end{equation}
and relax to the Boltzmann 
distribution~(\ref{eq: Master equations as a matrix equation e}) 
as $\tau\to\infty$. 
Here, the right-eigenvalues $\{\lambda^{\ }_{n}\}$ and right-eigenvectors 
$\{\psi^{(R;n)}_{\mathcal{C}}\}$ of the transition matrix 
$(W^{\ }_{\mathcal{C}\mathcal{C}'})$ control the relaxational 
time-dependence of the 
solutions~(\ref{eq: Solutions Master equations as a matrix equation}) 
to the Master equation~(\ref{eq: Master equations as a matrix equation}). 
Note that the spectral representation of the (not necessarily symmetric) 
matrix $(W^{\ }_{\mathcal{C}\mathcal{C}'})$ requires the 
introduction of left-eigenvectors $\{\psi^{(L;n)}_{\mathcal{C}}\}$ 
in addition to the right-eigenvectors and takes the form 
\begin{equation}
\label{eq: W decomposition}
  W^{\ }_{\mathcal{C}\mathcal{C}'} = 
      \sum_{n\in\mathbb{N}} \lambda^{\ }_{n} 
      \psi^{(R;n)}_{\mathcal{C}} \psi^{(L;n)}_{\mathcal{C}'}, 
\end{equation}
\begin{equation}  
  \delta^{\ }_{mn} = 
    \sum_{\mathcal{C}\in\mathcal{S}} 
      \psi^{(L;n)}_{\mathcal{C}} \psi^{(R;m)}_{\mathcal{C}}.
\end{equation}
Comparing Eq.~(\ref{eq: W decomposition}) with the 
definition~(\ref{eq: def tilde W and W b}), 
one can verify the one-to-one correspondence 
\begin{subequations}
\label{eq: 1to1 correspondence}
  \begin{equation}
    \lambda^{\ }_{n} = -\varepsilon^{\ }_{n},
  \end{equation}
  \begin{equation}
    \psi^{(R;n)}_{\mathcal{C}} = 
      e^{-K E^{\ }_{\mathcal{C}}/2} \times 
        \langle\mathcal{C}|\varepsilon^{\ }_{n}\rangle,
  \end{equation}
  \begin{equation}
    \psi^{(L;n)}_{\mathcal{C}} = 
      e^{+K E^{\ }_{\mathcal{C}}/2} \times
        \langle\varepsilon^{\ }_{n}|\mathcal{C}\rangle,
  \end{equation} 
\end{subequations}
with the eigenvectors and eigenvalues of the quantum Hamiltonian 
$\widehat{H}^{\ }_{SMF}$ in 
Eq.~(\ref{eq: notation for spectral decomposition HRK}), in the 
preferred basis. 
This fundamental property is a direct consequence of the SMF decomposition
of the Hamiltonian~(\ref{eq: Master equations as a matrix equation a}) and 
plays a key role in establishing the equivalence between classical and 
quantum correlation functions, proved by Henley in Ref.~\cite{Henley04}. 

We close the discussion on the classical stochastics induced by 
a quantum Hamiltonian that admits an SMF decomposition
by giving the explicit form of the 
transition matrix in the preferred basis. 
Substituting Eqs.~(\ref{eq: H RK positive def a},%
\ref{eq: H RK positive def b}) into the 
definitions~(\ref{eq: def tilde W and W}) we obtain 
%
%
\begin{equation}
\label{eq: W elements}
W^{\ }_{\mathcal{C}\mathcal{C}'} = \left\{ 
  \begin{array}{ll}
    w^{\ }_{\mathcal{C},\mathcal{C}'} 
      e^{-K (E^{\ }_{\mathcal{C}}-E^{\ }_{\mathcal{C}'})/2}, & 
        \textrm{if } (\mathcal{C},\mathcal{C}')\in\mathcal{L}, 
    \\
    &\\
    0, & \textrm{if } (\mathcal{C},\mathcal{C}')\notin\mathcal{L},
    \\
    &\\
    -\sum\limits_{\mathcal{C}''\in\mathcal{S}}
                   ^{\mathcal{C}''\neq\mathcal{C}} 
     W^{\ }_{\mathcal{C}''\mathcal{C}},
    &
    \textrm{if } \mathcal{C}=\mathcal{C}'.
  \end{array}
  \right. 
\end{equation}
Notice that the last line of 
Eq.~(\ref{eq: W elements}) is guaranteed to hold by 
conditions~(\ref{eq: Master equations as a matrix equation c}). 
For the purpose of performing numerical simulations of the classical system, 
it is worth noticing that the special cases of Metropolis and Glauber 
dynamics can be implemented by choosing 
\begin{subequations}
  \begin{equation}
  \label{Metrop_w}
    w^{\ }_{\mathcal{C},\mathcal{C}'} \propto 
      \max\left\{e^{-K (E^{\ }_{\mathcal{C}'}-E^{\ }_{\mathcal{C}})/2},
            e^{+K (E^{\ }_{\mathcal{C}'}-E^{\ }_{\mathcal{C}})/2}\right\}
  \end{equation}
and 
  \begin{equation}
  \label{Glaub_w}
    w^{\ }_{\mathcal{C},\mathcal{C}'} \propto 
      \frac{1}{e^{-K (E^{\ }_{\mathcal{C}'}-E^{\ }_{\mathcal{C}})/2} + 
               e^{+K (E^{\ }_{\mathcal{C}'}-E^{\ }_{\mathcal{C}})/2}}, 
  \end{equation}
\end{subequations}
respectively (recall that $w^{\ }_{\mathcal{C},\mathcal{C}'}$ is required to 
be symmetric upon exchanging $\mathcal{C}$ and $\mathcal{C}'$ by 
construction). 

We have shown in Sec.~\ref{sec: Generalized quantum dimer models} 
that the coefficients $w^{\ }_{\mathcal{C},\mathcal{C}'}$ remain free 
after relating the GS of a quantum Hamiltonian that admits an 
SMF decomposition to the 
partition function of a classical system in thermal equilibrium. 
We now see that the coefficients $w^{\ }_{\mathcal{C},\mathcal{C}'}$ play a 
crucial role when extending the relation between the quantum and classical 
systems to the excitation and relaxation spectra, respectively. 
We note that there are precisely as many independent parameters 
$w^{\ }_{\mathcal{C},\mathcal{C}'}$ 
as there are independent non-vanishing matrix elements in the transition 
matrix $(W^{\ }_{\mathcal{C},\mathcal{C}'})$. 
This fact has two consequences: 
\begin{itemize}
\item[(1)] 
If we equip the classical system~(\ref{eq: def Z(K)}) with some 
stochastics that can be written as the Master equation of the matrix 
type~(\ref{eq: Master equations as a matrix equation}), 
then there exists a unique set $\{w^{\ }_{\mathcal{C},\mathcal{C}'}\}$ 
that implements the quantum dynamics in some representation of 
the quantum Hamiltonian 
$(\mathcal{L},
  w^{\ }_{\mathcal{C},\mathcal{C}'},
  E^{\ }_{\mathcal{C}},
  K)
$ 
in the sense that 
relaxation and quantum normal modes are identical. 
Thus, the representation~(\ref{eq: H RK positive def}) 
for a quantum Hamiltonian is indeed 
the most general one that follows from using Henley's procedure 
in Ref.~\cite{Henley04} 
to construct a quantum Hamiltonian satisfying the RK conditions
(I-III) starting from a Master equation of the matrix 
type~(\ref{eq: Master equations as a matrix equation}). 
\item[(2)] 
Starting from a quantum Hamiltonian that admits an SMF decomposition~%
$(\mathcal{L},
  w^{\ }_{\mathcal{C},\mathcal{C}'},
  E^{\ }_{\mathcal{C}},
  K)
$, 
we have constructed a unique and generic classical system equipped with 
time stochastics through 
the Master equation~(\ref{eq: Master equations as a matrix equation}) 
whose relaxation rates are identical, up to an overall sign, 
to the energy eigenvalues of the quantum Hamiltonian
$(\mathcal{L},
  w^{\ }_{\mathcal{C},\mathcal{C}'},
  E^{\ }_{\mathcal{C}},
  K)
$. 
\end{itemize}
Combining (1) and (2), we have established a one-to-one correspondence 
between quantum Hamiltonians that admit SMF 
decompositions~(\ref{eq: H RK positive def}) 
and stochastic classical systems that can be represented by the 
Master equation of the matrix 
type~(\ref{eq: Master equations as a matrix equation}). 

\section{Two Theorems about the Stochastic Matrix Form decomposition}
\label{sec: Beyond RK Hamiltonians: The Stochastic Matrix Form}

In the two previous sections we have shown how one can 
deduce from the known examples 
of quantum Hamiltonians fine-tuned to their RK points 
a general class of matrix representations of quantum Hamiltonians 
that exhibit the same characteristic properties, and how 
these matrix representations are related to stochastic classical systems. 
In this construction, the positive semi-definite decomposition 
conditions~(\ref{eq: condition for positive definite}) and the integrability 
conditions~(\ref{eq: integrability conds}) play a crucial role. 
Both conditions depend not only on the choice of the preferred basis 
$\mathcal{B}=\{|\mathcal{C}\rangle, \; \mathcal{C}\in\mathcal{S}\}$ 
but also on the non-unique decomposition~(\ref{eq: H elements a}) 
of the diagonal matrix elements of the Hamiltonian $\widehat{H}$. 
We shall show in this section under what conditions there exists a unique 
decomposition of the diagonal elements of a quantum Hamiltonian $\widehat{H}$ 
in a suitable irreducible representation that satisfies the positive 
semi-definite decomposition 
conditions~(\ref{eq: condition for positive definite}) and the integrability 
conditions~(\ref{eq: integrability conds}). 

Also, we have shown in Sec.~\ref{sec: From quantum to classical dynamics} 
that, to any SMF representation of a quantum Hamiltonian $\widehat{H}$, 
in some preferred basis 
$\mathcal{B}=\{|\mathcal{C}\rangle, \; \mathcal{C}\in\mathcal{S}\}$, 
there corresponds a unique Master equation of the matrix type. 
We shall give below sufficient conditions under which 
a quantum Hamiltonian $\widehat{H}$, when construed as an abstract operator 
acting on a Hilbert space, 
admits different SMF representations and thus corresponds to different 
classical systems described by Master equations of the matrix type. 

\begin{thm}
\label{Theorem 1}
Given any real Hamiltonian $\widehat{H}$ defined on a finite dimensional 
Hilbert space $\mathcal{H}$ such that 
(i)~it has a vanishing ground state energy, 
(ii)~there exists a basis 
$\mathcal{B}=\{|\mathcal{C}\rangle, \; \mathcal{C}\in\mathcal{S}\}$ 
such that $\widehat{H}$ is represented by the irreducible matrix 
$(H^{\ }_{\mathcal{C}\mathcal{C}'})$, 
and 
(iii)~$H^{\ }_{\mathcal{C}\mathcal{C}'}\le 0$\; 
for $\mathcal{C}\ne \mathcal{C}'$, 
then $(H^{\ }_{\mathcal{C}\mathcal{C}'})$ is an SMF decomposable 
representation which is unique in every such basis $\mathcal{B}$, 
up to a permutation of the basis elements.
\end{thm}

An immediate corollary is that the RK conditions (I-III) in 
Sec.~\ref{sec: Introduction} are \textit{always} 
satisfied by a matrix that meets the hypotheses of Theorem~\ref{Theorem 1}. 
In this sense, there is nothing special about the known examples of quantum 
Hamiltonians tuned to their RK points 
with respect to any other value of the parameters, 
other than the fact that the SMF decomposition is promptly handed-in 
from the knowledge of the energies 
$\{E^{\ }_{\mathcal{C}},\; \mathcal{C}\in\mathcal{S}\}$ 
(through the ground state wavefunction) and the off-diagonal matrix elements 
$H^{\ }_{\mathcal{C}\mathcal{C}'}\le 0$\; 
for $\mathcal{C}\ne\mathcal{C}'$. 
As an example, we can apply Theorem~\ref{Theorem 1} to the quantum dimer 
problem on the square lattice~\cite{Kivelson87,Rokhsar88}. 
We recall that the quantum dimer Hamiltonian is parametrized by two 
coupling constants commonly denoted $v\in(-\infty,+\infty)$ and 
$t\in(0,\infty)$ for the diagonal and off-diagonal contributions to the 
quantum dimer Hamiltonian represented in the usual 
classical-configuration-state basis, respectively. 
The RK point $t=v$ is the one for which the ground state wave function can 
be constructed explicitly and shown to have a vanishing energy eigenvalue. 
However, since the quantum dimer Hamiltonian is real-valued and its 
off-diagonal elements are non-positive, with the restriction to limit the 
Hilbert space to one of the sectors in which the Hamiltonian is irreducible 
it is possible to apply Theorem~\ref{Theorem 1} for \textit{any value} of 
the dimensionless coupling $t/v$ up to a shift of the spectrum that ensures 
that the ground state has a vanishing eigenvalue. 
Hence, the quantum dimer Hamiltonian allows for an SMF 
decomposition in the classical dimer configuration basis for any value of 
$t/v$. In other words, it is always possible to write the quantum dimer 
Hamiltonian as a sum of $2\times2$ projectors in the 
classical-configuration-state basis, but the zero-energy GS becomes 
a superposition of classical configuration states with positive coefficients 
that are no longer required to be all equal. 
If so, the spectral properties of the quantum dimer model at any $t/v$ other 
than the RK point $t=v$ can also be recast as a problem in classical 
stochastic dynamics and thus amenable to classical simulations as opposed to 
quantum simulations. 
Of course, aside from the problem of solving for the ground state, 
it is likely that the classical stochastic dynamics cannot be implemented by 
local updates of the classical configurations induced by the SMF 
representation. 

An immediate generalization of Theorem~\ref{Theorem 1} can be achieved by 
considering complex Hamiltonians that satisfy hypotheses~(i) and~(ii) but 
fail to satisfy hypothesis~(iii), provided all the off-diagonal elements 
can be made real and non-positive via a gauge transformation, i.e., provided 
there exists a unitary diagonal basis transformation 
\begin{equation}
|\mathcal{C}\rangle\to
e^{i\phi^{\ }_{\mathcal{C}}}|\mathcal{C}\rangle,
\qquad \phi^{\ }_\mathcal{C}\in(0,2\pi)
\end{equation}
that makes all transformed off-diagonal matrix elements non-positive. 

On completely general grounds, one can formulate an even stronger theorem. 

\begin{thm}
\label{Theorem 2}
Any Hamiltonian $\widehat{H}$ defined on a finite dimensional Hilbert space 
$\mathcal{H}$ with a non-degenerate ground state whose eigenvalue is vanishing 
admits a continuous manifold of distinct block diagonal representations where 
each irreducible block is an independent SMF decomposition. 
\end{thm}

Theorem~\ref{Theorem 2} immediately implies that the correspondence between 
a quantum Hamiltonian $\widehat{H}$ satisfying the hypotheses of 
Theorem~\ref{Theorem 2}
and stochastic classical systems is one-to-(continuously) many. 
The assumption that the ground state has a vanishing eigenvalue 
is not essential since one can always achieve this by shifting 
rigidly the energy spectrum by the subtraction 
of the unit matrix multiplied by the ground state energy. 

We now turn to the proofs of Theorems~\ref{Theorem 1} and~\ref{Theorem 2}. 

\begin{pf*}{Proof (Theorem~\ref{Theorem 1})}
Let $(H^{\ }_{\mathcal{C}\mathcal{C}'})$ be the matrix representation 
of the quantum Hamiltonian $\widehat{H}$ that satisfies the hypotheses of 
Theorem~\ref{Theorem 1}. All the off-diagonal matrix elements of 
$-(H^{\ }_{\mathcal{C}\mathcal{C}'})$ are non-negative. 
By choosing a suitable positive constant $c$, all matrix elements of 
$-(H^{\prime}_{\mathcal{C}\mathcal{C}'}):= 
 -(H^{\ }_{\mathcal{C}\mathcal{C}'})
 +c\,(\delta^{\ }_{\mathcal{C}\mathcal{C}'})$ 
can be made non-negative. As such, the matrix 
$-(H^{\prime}_{\mathcal{C}\mathcal{C}'})$ 
obeys the hypotheses of the strong form of the Perron-Frobenius theorem 
from Ref.~\cite{Perron-Frobenius} 
(if the matrix $-(H^{\prime}_{\mathcal{C}\mathcal{C}'})$ were to be 
reducible, i.e., $(H^{\ }_{\mathcal{C}\mathcal{C}'})$ is reducible, 
it would obey only the hypotheses of the weak form of the 
Perron-Frobenius theorem from Ref.~\cite{Perron-Frobenius}. 
However, one can always work separately within each irreducible block). 
The application of the strong form of the Perron-Frobenius theorem to the 
matrix $-(H^{\prime}_{\mathcal{C}\mathcal{C}'})$ 
implies that its eigenvector $\Psi^{\ }_{0}$ with the largest eigenvalue 
is a (non-degenerate) strictly positive eigenvector, i.e., 
$\Psi^{\ }_{0,\mathcal{C}}>0,\;\forall\,\mathcal{C}$. 
Since $-(H^{\prime}_{\mathcal{C}\mathcal{C}'})$ and 
$(H^{\ }_{\mathcal{C}\mathcal{C}'})$ share the same eigenvectors, 
the strictly positive eigenvector ${\Psi}^{\ }_{0}$ is also the ground 
state with \textit{vanishing} eigenvalue of 
$(H^{\ }_{\mathcal{C}\mathcal{C}'})$. 

We can then identify a set of Boltzmann weights from the strictly 
positive coefficients of $\Psi^{\ }_{0}$, assumed normalized to one, 
and we can read out the classical energies 
$\{KE^{\ }_{\mathcal{C}}:=-2\ln\left(\Psi^{\ }_{0,\mathcal{C}}\right), 
\;\mathcal{C}\in\mathcal{S}\}$ from these Boltzmann weights, 
apart from an irrelevant shift of the energies 
$\{E^{\ }_{\mathcal{C}},\; \mathcal{C}\in\mathcal{S}\}$ by 
the constant $\ln{[Z(K)]}/K$. 
Using these classical energies 
$\{E^{\ }_{\mathcal{C}},\; \mathcal{C}\in\mathcal{S}\}$ 
we can perform a similarity transformation on the matrix 
$-({H}^{\ }_{\mathcal{C}\mathcal{C}'})$ 
as was done in Sec.~\ref{sec: From quantum to classical dynamics} 
and get the matrix $({W}^{\ }_{\mathcal{C}\mathcal{C}'})$ 
such that $({W}^{\ }_{\mathcal{C}\mathcal{C}'})$ has the left-eigenvector 
$(1,1,...,1)$ with \textit{vanishing} left-eigenvalue. As the off-diagonal 
elements of the matrix $-(H^{\ }_{\mathcal{C}\mathcal{C}'})$ are non-negative 
and the Boltzmann weights are positive (${\Psi}^{\ }_{0}$ is positive), 
the off-diagonal elements of $({W}^{\ }_{\mathcal{C}\mathcal{C}'})$ are 
also non-negative. 
Moreover, the \textit{vanishing} left-eigenvalue of $(1,1,...,1)$ 
ensures that the diagonal elements of $({W}^{\ }_{\mathcal{C}\mathcal{C}'})$ 
are minus the sum of the off-diagonal elements in the column 
to which they belong. Put together, this means that elements of 
$({W}^{\ }_{\mathcal{C}\mathcal{C}'})$ 
have a transition probability interpretation. If so we can make use of the 
fact that to each Master equation of the matrix type there corresponds a 
unique matrix obeying the SMF decomposition to conclude that the matrix 
$(H^{\ }_{\mathcal{C}\mathcal{C}'})$ realizes an RK point.

Since the components of the wave function $\Psi^{\ }_{0}$ 
are unique up to a permutation of the elements in the set $\mathcal{S}$, 
so are the Boltzmann weights, and so is the SMF~(\ref{eq: H RK positive def}) 
of the matrix $({H}^{\ }_{\mathcal{C}\mathcal{C}'})$. 
\qed
\end{pf*}

\begin{pf*}{Proof (Theorem~\ref{Theorem 2})}
Let $\widehat{H}$ be a quantum Hamiltonian that satisfies the hypotheses of 
Theorem~\ref{Theorem 2}. We are going to construct a continuous manifold of 
matrix representations of $\widehat{H}$ that satisfy the 
hypotheses of Theorem~\ref{Theorem 1}. 

We choose a basis of $\mathcal{H}$ that diagonalizes $\widehat{H}$, i.e., 
$\widehat{H}$ admits the matrix representation 
\begin{equation}
H^{\ }_{ij}=
\varepsilon^{\ }_{i}\delta^{\ }_{ij},
\qquad
i,j=0,\cdots,|\mathcal{S}|-1,
\end{equation}
such that all its eigenvalues are ordered according to 
\begin{equation}
\varepsilon^{\ }_{0}<
\varepsilon^{\ }_{1}\leq
\cdots\leq
\varepsilon^{\ }_{|\mathcal{S}|-1}.
\label{eq: ordering bar varepsilon}
\end{equation}
By hypothesis, $\varepsilon^{\ }_{0}=0$. 
Let $\widehat{A}$ be any operator defined on $\mathcal{H}$ such that it can 
be represented by a real-valued antisymmetric matrix whose matrix elements 
\begin{equation}
A^{\ }_{ij}=-A^{\ }_{ji}
\label{eq: A as}
\end{equation}
are all infinitesimal and strictly positive if $i<j$, 
\begin{equation}
i<j\Longrightarrow A^{\ }_{ij}>0.
\label{eq: Aij negative i i<j}
\end{equation}
Define 
\begin{equation}
\widehat{H}^{(A)}:=
e^{-\widehat{A}}\widehat{H}e^{+\widehat{A}},
\qquad
\delta^{(A)}\widehat{H}:= 
e^{-\widehat{A}}\widehat{H}e^{+\widehat{A}}-\widehat{H}.
\end{equation}
The Hamiltonian $\widehat{H}^{(A)}$ shares the same eigenvalues 
as $\widehat{H}$ but is not diagonal in the basis that diagonalizes 
$\widehat{H}$. 
Because $\widehat{A}$ is infinitesimal we can write 
\begin{equation}
\delta^{(A)}\widehat{H}=[\widehat{H},\widehat{A}].
\end{equation}
The matrix elements of $\delta^{(A)}\widehat{H}$ 
in the basis that diagonalizes $\widehat{H}$ are 
\begin{equation}
(\delta^{(A)} H)^{\ }_{ij}= 
(\varepsilon^{\ }_{i}-\varepsilon^{\ }_{j}) A^{\ }_{ij}=
(\delta^{(A)} H)^{\ }_{ji}.
\end{equation}
In view of Eqs.~(\ref{eq: ordering bar varepsilon}), 
(\ref{eq: A as}), and~(\ref{eq: Aij negative i i<j}), 
\begin{equation}
(\delta^{(A)} H)^{\ }_{ij}\leq 0
\end{equation}
where the equality holds if there are degenerate eigenvalues or if $i=j$. 
We conclude that the matrix representation 
\begin{equation}
H^{(A)}_{ij}=
H^{\ }_{ij}
+
(\delta^{(A)} H)^{\ }_{ij}
\end{equation} 
obeys the hypotheses of Theorem~\ref{Theorem 1} at least within every 
irreducible block, once the basis vectors are premuted in such a way as to 
make the representation $H^{(A)}_{ij}$ block diagonal. Note that if none 
of the eigenstates has degeneracy larger than or equal to 
$\vert S \vert/2$, irreducibility is ensured. 
\qed
\end{pf*}

\section{Examples}
\label{sec: Examples}

Before reviewing a few examples of quantum Hamiltonians 
that admit SMF representations, 
it is worth mentioning that the notation chosen in 
Eqs.~(\ref{eq: H RK positive def}), although the most general, can usually 
be further simplified depending on the specific details of the system. 
In fact, independently of whether the starting point is a classical or 
quantum system, the set of changes in phase space $(\mathcal{C},\mathcal{C'})$ 
produced by the classical stochastics or by the kinetic energy operator 
can often (although not always) be interpreted as local (on some lattice) 
rearrangements $\ell$ of degrees of freedom 
(e.g., spin flips or exchanges, plaquette dimer operations, loop updates). 
Under the condition that the energy change is local, all the relevant 
physics in a pair $(\mathcal{C},\mathcal{C'})$ depends only on a neighborhood 
of the local rearrangement $\ell$ that connects $\mathcal{C}$ to 
$\mathcal{C}'$ 
(i.e., $E^{\ }_{\mathcal{C}'}-E^{\ }_{\mathcal{C}}=\varepsilon^{\ }_{\ell}$ 
and 
$w^{\ }_{\mathcal{C},\mathcal{C}'}=w^{\ }_{\ell}$). 
Then, the summation over pairs $(\mathcal{C},\mathcal{C}')\in\mathcal{L}$ can 
be reduced to a summation over local rearrangements $\ell\in\mathcal{L}^r$, 
and the quantum Hamiltonian at its RK point assumes a simpler, 
more intuitive form. Practical examples of this simplification 
are given hereafter. 

We will first provide two known examples of Hamiltonians fine-tuned to their
RK points and show how they relate to our general formalism. 
We will also suggest possible generalizations 
of these models that arise naturally from our results. 
Then, we will provide an example for possible practical applications of the 
results presented in Sections~\ref{sec: Generalized quantum dimer models} 
and~\ref{sec: From quantum to classical dynamics}. 

\subsection{The triangular lattice quantum dimer model} 
\label{sec: quantum dimers on triangular lattice}

The quantum dimer model on the triangular lattice is defined on 
the separable Hilbert space 
spanned by the preferred, orthonormal basis 
$\mathcal{B}=\{|\mathcal{C}\rangle,\, \mathcal{C}\in\mathcal{S}\}$, 
where $\mathcal{C}$ stands for any of the 
possible close-packed, hardcore dimer coverings of the triangular lattice. 
This Hilbert space is not connected under the time-evolution operator 
generated by 
\begin{equation} 
\begin{split}
  \widehat{H} &:= 
    - t \widehat{T} + v \widehat{V} 
  \\
  &\equiv 
  \sum_{i=1}^{N^{\ }_p}\sum_{\alpha=1}^3   
    \left[ 
      - t 
      \left( 
        |\input{tridimer1.tex}\rangle \langle\input{tridimer2.tex}| + 
        H.c. 
      \right) 
    + v 
      \left( 
        |\input{tridimer1.tex}\rangle \langle\input{tridimer1.tex}| + 
	|\input{tridimer2.tex}\rangle \langle\input{tridimer2.tex}| 
      \right)
  \right], 
\end{split}
\label{eq: triham}
\end{equation}
where the sum on $i$ runs over all of the $N^{\ }_p$ plaquettes 
\input{tridimerplaq.tex} making up the triangular lattice 
while the sum on $\alpha$ runs over all three even permutations 
of the three sublattices making up the triangular lattice~\cite{Moessner01}. 
We call a plaquette occupied by a pair of parallel dimers flippable. 
This Hamiltonian is known to exhibit an RK point at 
$v=t\:(=1)$~\cite{Moessner01}. At this RK point, one of the possible GS 
is the equal-weight superposition of all the elements 
of the preferred basis~\cite{Moessner01}. 
Hence, the normalization of this GS is nothing but 
the purely combinatorial, classical problem of counting all the 
arrangements of hardcore and close-packed dimers on a triangular lattice. 

We are now going to show how Eq.~(\ref{eq: triham}) with $v=t=1$ 
relates to Eq.~(\ref{eq: H RK positive def}) with $K=0$, 
$w^{\ }_{\mathcal{C},\mathcal{C}'}=1$, and $\mathcal{S}$ defined as in 
the last paragraph. First, we need to construct the set 
$\mathcal{L}\subset\mathcal{S}\times\mathcal{S}$. 
To this end, we define $\mathcal{L}$ to be the set of all pairs 
of configurations in $\mathcal{S}\times\mathcal{S}$ 
that are mapped one onto the other by a single flippable plaquette update, 
i.e., $\mathcal{C}$ differs from $\mathcal{C}'$ 
by one and only one flippable plaquette, 
say $\input{tridimer1.tex}$ for $\mathcal{C}$ 
and $\input{tridimer2.tex}$ for $\mathcal{C}'$, 
whenever $(\mathcal{C},\mathcal{C}')\in\mathcal{L}$. 
Equipped with $\mathcal{L}$, we can then consider 
Eq.~(\ref{eq: H RK positive def}) with $K=0$ and 
$w^{\ }_{\mathcal{C},\mathcal{C}'}=1$, and rearrange it using the fact 
that the operators 
$|\mathcal{C}\rangle\langle\mathcal{C}|$ 
and 
$|\mathcal{C}'\rangle\langle\mathcal{C}|$ 
both enter Eq.~(\ref{eq: H RK positive def}) 
with the same weights. For any given plaquette \input{tridimerplaq.tex} 
and up to an even permutation of the three sublattices of the triangular 
lattice, it is possible to perform the two sums~\cite{note: ell in C} 
\begin{subequations} 
  \begin{equation}
  \label{eq: resummed op 1}
  |\input{tridimer1.tex}\rangle\langle\input{tridimer1.tex}| := 
    \sum_{\input{tridimer1_sm.tex}\in\mathcal{C}} 
      |\mathcal{C}\rangle\langle\mathcal{C}|, 
  \end{equation}
and 
  \begin{equation}
  \label{eq: resummed op 2} 
  |\input{tridimer2.tex}\rangle\langle\input{tridimer1.tex}| := 
    \sum_{\input{tridimer1_sm.tex}\in\mathcal{C}}
        ^{\stackrel{
	    \!\!\!\!\input{tridimer1_sm.tex} \rightarrow\input{tridimer2_sm.tex}}
	    {\mathcal{C}\;\longrightarrow\;\mathcal{C}'}} 
      |\mathcal{C}'\rangle\langle\mathcal{C}|. 
  \end{equation}
\end{subequations} 
Here, $\input{tridimer1.tex}\in\mathcal{C}$ is a shortcut for 
$\{\mathcal{C},\textrm{ s.t. }\input{tridimer1.tex}\in\mathcal{C}\}$, while 
$\stackrel{\!\!\!\!\input{tridimer1_sm.tex}\rightarrow\input{tridimer2_sm.tex}}
          {\mathcal{C}\:\longrightarrow\:\mathcal{C}'}$ stands for 
$\left\{\mathcal{C}', \textrm{ s.t. } \mathcal{C}\right.$ transforms into 
$\mathcal{C}'$ upon the local rearrangement 
$\left. \input{tridimer1.tex}\rightarrow\input{tridimer2.tex}\right\}$. 
We have thus recovered Eq.~(\ref{eq: triham}) starting from 
Eq.~(\ref{eq: H RK positive def}) 
with $K=0$ and $w^{\ }_{\mathcal{C},\mathcal{C}'}=1$. 
Of course there is nothing sacred with these two conditions. 
We might as well relax them without spoiling the RK properties (I-III) of our 
quantum Hamiltonian which now takes the more general 
form~\cite{note: RK dimer conditions} 
\begin{equation}
\begin{split} 
\widehat{H}^{\ }_{SMF} = \frac{1}{2} 
 \sum_{\input{tridimer1.tex}\in\mathcal{L}^r}\!\!
  w^{\ }_{\input{tridimer1_sm.tex}} 
  \left[\vphantom{\sum}
- 
    \left(
      |\input{tridimer2.tex}\rangle\langle\input{tridimer1.tex}| + 
      H.c. 
    \right) 
  \right. 
\qquad \qquad \qquad \qquad \qquad
\\ 
\qquad \qquad \qquad \qquad \qquad
  \left.    
+ 
    e^{-K \varepsilon^{\ }_{\input{tridimer1_sm.tex}} /2} 
      |\input{tridimer1.tex}\rangle\langle\input{tridimer1.tex}|
+ 
    e^{+K \varepsilon^{\ }_{\input{tridimer1_sm.tex}} /2} 
      |\input{tridimer2.tex}\rangle\langle\input{tridimer2.tex}|
  \vphantom{\sum}\right] 
\end{split}
\label{eq: def triangular RK general}  
\end{equation} 
whereby we have replaced the double sum in Eq.~(\ref{eq: triham}) with 
the single sum labeled by ${\input{tridimer1.tex}\in\mathcal{L}^r}$, where 
$\mathcal{L}^r$ is the set of all possible plaquettes with two parallel 
dimers (i.e., $\input{tridimer1.tex}$ or $\input{tridimer2.tex}$ for all 
positions $i$ and directions $\alpha$). The reasons for such a change in 
notation will be clear from the examples presented in the following Sections. 
Notice that a factor of $1/2$ is now required to avoid double counting due to 
the distinction between the two dimer coverings $\input{tridimer1.tex}$ and 
$\input{tridimer2.tex}$ of the same plaquette. 
The Hamiltonian $\widehat{H}^{\ }_{SMF}$ 
differs from the quantum Hamiltonian~(\ref{eq: triham}) fine-tuned to its 
RK point $t=v$ by the presence of the classical 
interaction between the dimers in the form of the configuration energy 
$E^{\ }_{\mathcal{C}}$, 
whereby 
$E^{\ }_{\mathcal{C}'}-E^{\ }_{\mathcal{C}} \equiv 
\varepsilon^{\ }_{\input{tridimer1_sm.tex}}$%
~\cite{note: decorated rearrangements}. 
One of the possible GS of Eq.~(\ref{eq: def triangular RK general}) 
has the classical partition function 
\begin{equation}
Z(K):=
\sum_{\mathcal{C}\in\mathcal{S}}e^{-KE^{\ }_{\mathcal{C}}}
\end{equation}
for normalization. Hence, the phase diagram of this 
classical partition function along the reduced temperature 
axis $K$ partly controls the zero-temperature phase diagram 
of the quantum Hamiltonian~(\ref{eq: def triangular RK general}) 
as a function of $K\in\mathbb{R}$. 
Finally, the energy spectrum can be tuned by the coupling constants 
$w^{\ }_{\mathcal{C},\mathcal{C}'} \equiv w^{\ }_{\input{tridimer1_sm.tex}}$. 

\subsection{The quantum eight-vertex model} 
\label{sec: quantum RK 8vertex}

\begin{figure}[ht]
\centering
\includegraphics[width=0.98\columnwidth]{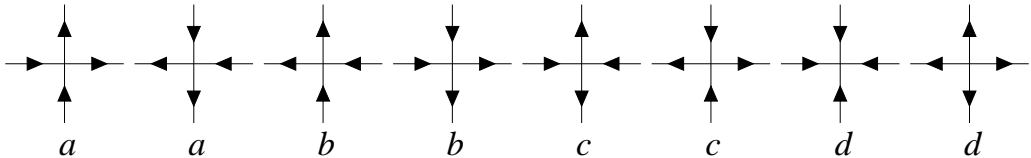} 
\caption{
\label{fig: 8 vertices}
The allowed eight type of vertices with the corresponding real-valued 
numbers $a$, $b$, $c$, and $d$. 
}
\end{figure}

We turn our attention to the quantum eight-vertex model at its RK point, 
as discussed by Ardonne, Fendley, and Fradkin~\cite{Ardonne04}. 
The classical configuration space $\mathcal{S}$ consists of 
all possible arrangements of vertices of the eight types shown 
in Fig.~\ref{fig: 8 vertices} so as to form a square lattice. 
The Boltzmann weight of a configuration $\mathcal{C}\in\mathcal{S}$ 
is obtained by taking the product of the positive numbers (fugacities) 
$a^2$, $b^2$, $c^2$, and $d^2$ associated to all the vertices appearing in 
$\mathcal{C}$, according to the pairings implied by Fig.~\ref{fig: 8 vertices}. 
We demand that the partition function respects the symmetries of the 
square lattice, in which case we must impose the condition $a=b$. 
For simplicity we set $a=b=1$. This gives the classical partition function 
\begin{equation}
\label{eq: def partition fct for 8vm}
Z(c^2,d^2) := \sum_{\mathcal{C}\in\mathcal{S}} 
  c^{2 N^{(c)}_{\mathcal{C}}} d^{2 N^{(d)}_{\mathcal{C}}}
\end{equation} 
where $N^{(c)}_{\mathcal{C}}$ is the total number 
of vertices of type $c$, say, in configuration $\mathcal{C}$. 
The separable Hilbert space $\mathcal{H}$ is the span of the preferred, 
orthonormal basis 
$\mathcal{B}=\{|\mathcal{C}\rangle,\;\mathcal{C}\in\mathcal{S}\}$. 
In this basis, the quantum eight-vertex Hamiltonian fine-tuned 
to its RK point is represented by 
\begin{subequations}
\label{eq: 8vertex ham}
  \begin{equation}
    \widehat{H}^{\ }_{RK} := \sum_i w^{\ }_i \widehat{Q}^{\ }_i, 
  \end{equation}
  \begin{equation}
    \widehat{Q}^{\ }_i := 
      \left( 
        \begin{array}{cc}
          \!\!\!v^{\ }_{i}& 
          -1 \\
          -1\;\;\;\; & 
           1/v^{\ }_{i}
        \end{array}
      \right), 
  \end{equation}
  \begin{equation} 
  \label{eq: n_i}
    \vphantom{\mathop{\sum}^{X}}   
    v^{\ }_{i}:= 
    c^{\,{\overline{n}}^{(c)}_{i}-{n}^{(c)}_{i}}\,
    d^{\,{\overline{n}}^{(d)}_{i}-{n}^{(d)}_{i}}. 
  \end{equation}
\end{subequations}
Here, the index $i$, which runs over all the sites of the square lattice, 
labels all the elementary plaquettes of the square lattice 
(square plaquettes) through their lower-left corners, 
the coupling constant $w^{\ }_i$ is a positive weight, 
and the $2\times2$ matrix $\widehat{Q}^{\ }_i\propto \widehat{Q}^{2 }_i$ 
relates any configuration $\mathcal{C}$ to the (unique) 
configuration $\overline{\mathcal{C}}$ obtained from $\mathcal{C}$ by 
reversing all the arrows on the edges of plaquette $i$%
~\cite{note: 2x2 RK matrix}. 
The positive integers ${n}^{(c)}_{i}$ and ${\overline{n}}^{(c)}_{i}$ 
count the numbers of vertices of type $c$, say, that appear on 
plaquette $i$ in configuration $\mathcal{C}$ and 
$\overline{\mathcal{C}}$, respectively. 

The Hamiltonian~(\ref{eq: 8vertex ham}) has the normalized GS wavefunction 
\begin{equation}
|\Psi^{\ }_{RK}\rangle = \frac{1}{\sqrt{Z(c^2,d^2)}} 
  \sum_{\mathcal{C}\in\mathcal{S}} 
    c^{N^{(c)}_{\mathcal{C}}} d^{N^{(d)}_{\mathcal{C}}} |\mathcal{C}\rangle 
\end{equation}
which is non-degenerate in any of its four topological sectors
if periodic boundary conditions are imposed
owing to the fact that all configurations are flippable~\cite{Ardonne04}. 
The coupling space of the quantum eight-vertex model fine-tuned
to its RK point is 
two-dimensional and is parametrized by $c$ and $d$, which can be 
taken to be positive without loss of generality. In this parameter space 
the quantum eight-vertex model fine-tuned to its RK point 
exhibits quantum phase transitions between a confining ordered phase, 
with broken $\mathbb{Z}^{\ }_2$ symmetry, 
and a deconfining disordered phase, with unbroken $\mathbb{Z}^{\ }_2$ 
symmetry~\cite{Ardonne04}. 
The lines of critical points at the boundaries between those 
two phases correspond to the classical six-vertex model case and its 
dual~\cite{Ardonne04}. 

\begin{figure}[ht]
\centering
\includegraphics[width=0.7\columnwidth]{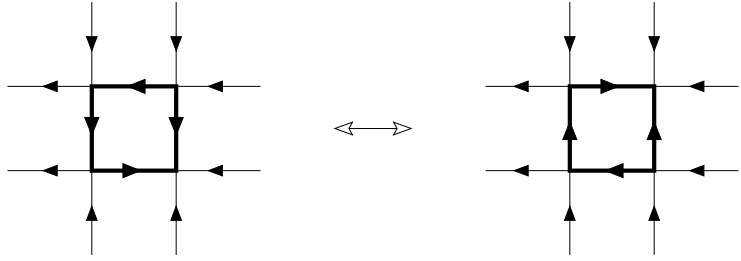}
\caption{
\label{fig: ex of local rearrangement}
Example of a \textit{decorated and oriented} plaquette in the eight-vertex 
model, and of the corresponding local rearrangement generated by the kinetic 
term in Eq.~(\ref{eq: 8vertex ham}). Notice that only the directions of the 
arrows along the square plaquette (thick line) get reversed. 
}
\end{figure}

In order to recast Eq.~(\ref{eq: 8vertex ham}) into the formalism of the 
past sections, observe that the potential energy term acts multiplicatively 
on a basis state $|\mathcal{C}\rangle$ with the proportionality constant 
given by 
\begin{equation}
  \sum_i w^{\ }_i \, c^{\overline{n}^{(c)}_i-n^{(c)}_i} 
                     d^{\overline{n}^{(d)}_i-n^{(d)}_i}
\end{equation}
The kinetic term acts instead on a state $|\mathcal{C}\rangle$ by mapping it 
to a state $|\overline{\mathcal{C}}\rangle$ (with coefficient $-1$) such that 
configurations $\mathcal{C}$ and $\overline{\mathcal{C}}$ differ by a single 
plaquette flip. Let us now define the set of local rearrangements needed to 
construct the reduced Hamiltonian fine-tuned to its RK point 
as the set of plaquette flips (i.e., 
reversal of the arrows on the edges of a plaquette) for all sites $i$ in the 
square lattice, and all the possible initial arrow orientations along the 
edges. An example of such plaquette rearrangements is given in 
Fig.~\ref{fig: ex of local rearrangement}. 
For reasons that will become clear in a moment, let us label these plaquette 
rearrangements by the set $\mathcal{L}^r$ of all 
\textit{decorated and oriented} plaquettes $\ell$, defined by the site $i$ 
of their lower-left corner, by the orientations of the arrows along the 
edges of the plaquette, and by the orientations of the arrows on all eight 
dangling bonds that connect the sites of the plaquette to 
(nearest-neighboring) sites not belonging to the plaquette 
(see Fig.~\ref{fig: ex of local rearrangement}). 
We will also denote by $\overline{\ell}$ the (unique) 
\textit{decorated and oriented} plaquette that is obtained from $\ell$ by 
reversing all the four arrows along the edges of the plaquette. 
Notice that the number of elements in $\mathcal{L}^r$ is $16$ times larger 
than the total number of simple \textit{oriented} plaquettes (i.e., where 
only the directions of the arrows along the edges are specified). 
This redundancy, although superfluous for the kinetic operator alone, 
plays a crucial role when writing the potential energy term in the reduced 
form. We can now reinterpret Eq.~(\ref{eq: 8vertex ham}) as 
\begin{subequations}
\begin{equation}
\label{eq: RK H 8vertex}
\widehat{H}^{\ }_{SMF} = \frac{1}{2} 
  \sum_{\ell\in\mathcal{L}^r} w^{\ }_{\ell} 
  \left[
    e^{-K \varepsilon^{\ }_{\ell} /2} \widehat{P}^{\ }_{\ell} + 
    e^{+K \varepsilon^{\ }_{\ell} /2} \widehat{P}^{\ }_{\overline{\ell}} - 
    \left(
      \widehat{T}^{\ }_{\ell,\overline{\ell}} + 
      \widehat{T}^{\ }_{\overline{\ell},\ell}
    \right)
  \right], 
\end{equation}
where~\cite{note: ell in C} 
\begin{equation} 
\label{eq: P T 8vertex} 
\widehat{P}^{\ }_{\ell} |\mathcal{C}\rangle = \left\{ 
  \begin{array}{ll}
    |\mathcal{C}\rangle, & \textrm{if } \ell\in\mathcal{C}, \\
    &\\ 
    0,                   & \textrm{otherwise}, 
  \end{array}
\right. 
\end{equation}
and
\begin{equation}
\widehat{T}^{\ }_{\ell,\overline{\ell}} |\mathcal{C}\rangle= \left\{ 
  \begin{array}{ll} 
    |\overline{\mathcal{C}}\rangle, & \textrm{if } \ell\in\mathcal{C}, \\ 
    &\\
    0,                              & \textrm{otherwise}. 
  \end{array}
\right. 
\end{equation}
\end{subequations}
Here, $w^{\ }_{\ell}$ is positive, $K$ is real-valued, 
$K \varepsilon^{\ }_{\ell} /2 = 
-(\overline{n}^{(c)}_{\ell}-n^{(c)}_{\ell})\ln(c) 
-(\overline{n}^{(d)}_{\ell}-n^{(d)}_{\ell})\ln(d)$%
~\cite{note: q8v integrability,note: ell vs i}, 
and $\overline{\mathcal{C}}$ is the (unique) configuration that is obtained 
from $\mathcal{C}$ after performing the plaquette flip 
$\ell\rightarrow\overline{\ell}$. 
We stress that the role of the label $\ell\in\mathcal{L}^{r}$ 
in Eq.~(\ref{eq: RK H 8vertex}) is two-fold. This label must specify the pair 
of classical configurations that ``resonate'' through a quantum 
tunneling process. This label must also carry the information needed to 
construct the difference between the classical configuration energies of the 
two classical configurations that ``resonate''. For the eight-vortex model 
with no classical interactions between the vertices, $\ell$ is obtained 
by decorating the oriented plaquettes with the additional information on the
orientations of the dangling bonds. 

\begin{figure}[ht]
\centering
\includegraphics[width=0.8\columnwidth]{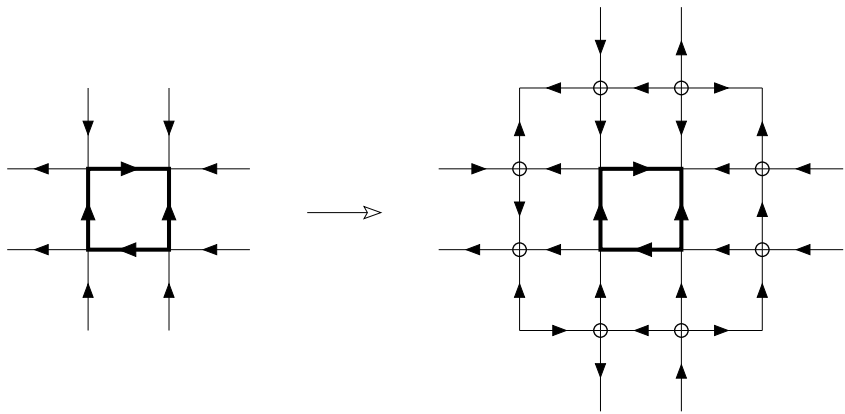} 
\caption{
\label{fig: q8v decor}
Example of the change in the definition of a \textit{decorated and oriented} 
loop $\ell$ required by the presence of nearest-neighbor interactions between 
the vertices. The eight nearest-neighboring vertices are highlighted with 
circles. 
}
\end{figure}

We have reinterpreted the quantum eight-vertex model fine-tuned to its 
RK point, introduced in Ref.~\cite{Ardonne04}, using the formalism presented 
in the past sections through Eq.~(\ref{eq: RK H 8vertex}). 
It is now straightforward to generalize Eq.~(\ref{eq: RK H 8vertex}), 
e.g., by adding interactions between the vertices as well as external 
fields. All this can easily be accommodated at the classical level, 
and thus at the quantum level as well, through a redefinition of 
$\varepsilon^{\ }_{\ell}$, and possibly of $\mathcal{L}^{r}$. 
Notice in fact that the definition of \textit{decorated and oriented} 
plaquettes $\mathcal{L}^{r}$ given above, while sufficient to describe an 
applied field or chemical potential on the vertices (as in the original case 
by Ardonne~\textit{et al.}), needs to be modified if interactions between 
vertices are introduced. For example, in case of nearest-neighbor 
interactions one has to encode in the definition of a 
\textit{decorated and oriented} loop $\ell$ the additional information about 
the types of nearest-neighboring vertices with respect to the ones belonging 
to the square plaquette, as illustrated in Fig.~\ref{fig: q8v decor}. 
In the presence of interactions between vertices, the coupling 
constant $K$ can be interpreted as the inverse temperature of the 
classical problem defined by the GS. 
In addition to changing 
$\mathcal{L}^{r}$ 
by extending the range of the decoration of an oriented plaquette, 
local rearrangements allowing moves in classical phase space 
different than single plaquette flips can be introduced. 
Any combination of these generalizations can change the 
zero-temperature phase diagram of the quantum model. 
Finally, we stress that the choice for the parameters $w^{\ }_{\ell}$ 
can affect in a crucial way the classical stochastics (excitation spectrum) 
of the classical (quantum) system. 

\subsection{The quantum three-coloring model} 
\label{sec: three coloring}

So far we have illustrated how the formalism of the past sections reproduces 
quantum Hamiltonians fine-tuned to their RK points
that have already been discussed in the 
literature, and how they can be generalized without spoiling 
the RK conditions (I-III). 
We devote the following discussion to the construction 
of a quantum Hamiltonian from an SMF decomposition whose 
classical counterpart exhibits dynamical glassiness. 

Consider the classical three-coloring model on the honeycomb lattice 
with nearest-neighbor interactions%
~\cite{Baxter70,Difrancesco94,Cirillo96,Castelnovo04}. 
The model consists of classical degrees of freedom living on the 
nearest-neighbor bonds of a honeycomb lattice. 
Periodic boundary conditions are assumed. Any of these classical degrees of 
freedom can assume three different values or colors (say, A, B, and C). 
The system is also subjected to the hard constraint that each and every color 
emanates from any site of the honeycomb lattice. 
The classical configuration space is denoted by $\mathcal{S}$. 
To each coloring of the honeycomb lattice $\mathcal{C}\in\mathcal{S}$ 
one can associate an Ising spin configuration with the Ising spins defined on 
the sites of the honeycomb lattice, according to the rule that a site is 
occupied by an up (down) Ising spin if the parity of the sequence of colors 
emanating from this site is even (odd) when encircling the site 
counter-clockwise, say. 
The classical energy of any coloring $\mathcal{C}\sim\{\sigma^{\mathrm{z}}_i\}$ 
of the honeycomb lattice is taken to be the Heisenberg energy 
\begin{subequations}
\label{def: 3color model}
  \begin{equation}
  \label{eq: E_c}
    E^{\ }_{\mathcal{C}} = -J \sum_{\langle i,j \rangle} 
      \sigma^{\mathrm{z}}_i \sigma^{\mathrm{z}}_j. 
  \end{equation}
Here,  $\langle i,j \rangle$ denotes an oriented pair of nearest-neighbor 
sites on the honeycomb lattice and $J\in\mathbb{R}$ is the spin stiffness. 
The classical partition function of the three-coloring model is then given 
by 
  \begin{equation}
  \label{def: 3color model Z(K)}
    Z(K) = \sum_{\mathcal{C}\in\mathcal{S}}
      e^{-KE^{\ }_{\mathcal{C}}}.
  \end{equation}
\end{subequations}
The color constraint becomes the requirement that the sum over all the Ising 
spins around any elementary plaquette of the honeycomb lattice, a hexagon, 
be $\pm6,0$ in the Ising spin representation. We refer the reader to 
Ref.~\cite{Castelnovo04} and references therein 
for a detailed analysis of the properties 
on the three-coloring model~(\ref{def: 3color model}) 
in and out of thermal equilibrium. 
Suffice here to say that, in addition to the usual Ising ordered phases 
(N\'eel and ferromagnetic), there exists a large-temperature critical phase 
for $0 \leq T/J \leq (T/J)_{cr}$ in thermal 
equilibrium~\cite{Kondev96,Castelnovo04}. 
This critical phase is expected to terminate in a first-order phase transition 
to the fully-magnetized ferromagnetic state for sufficiently positive $J$. 
The ferromagnetic phase has very interesting properties out of thermal 
equilibrium. At and beyond this first order phase transition, the system 
encounters a dynamical obstruction to equilibration 
(classical dynamical glassiness) that gives origin to a supercooled liquid phase. 
The supercooled liquid phase freezes at a temperature $(T/J)<(T/J)_{cr}$ into 
a polycrystallized phase with no interstitial liquid left. 
In view of the intimate connection between classical systems out of 
equilibrium and quantum Hamiltonians that admit SMF representations, 
it is then natural to ask if there is a signature of the classical dynamical 
glassiness in such quantum systems. 
The detailed answer to this question is beyond the scope of this paper. 
Below, we limit ourselves to the construction of an SMF decomposition of 
a quantum Hamiltonian that has the three-coloring model in thermal 
equilibrium emerging from its GS. 

\begin{figure}[ht]
\centering
\includegraphics[width=1.00\columnwidth]{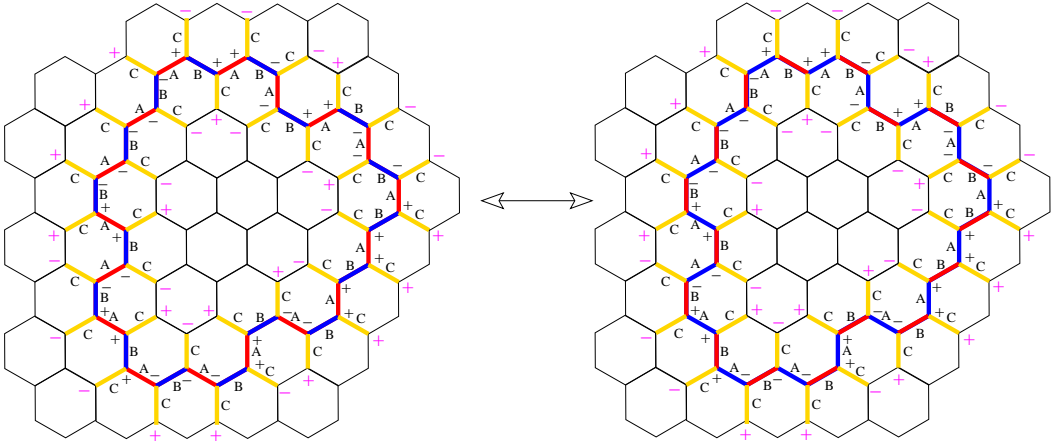} 
\caption{
\label{fig: 3col decor}
Example of a \textit{decorated} loop in the three-coloring model, as 
required by the presence of nearest-neighbor interactions between the 
spins, and of the corresponding rearrangement. 
Notice the information encoded in the loop: 
(1) the backbone structure, given by the sites of all the vertices visited 
by the AB loop; 
(2) the two-color sequence covering, given by the specific labeling of all 
the bonds belonging to the loop; and 
(3) the values of the nearest-neighboring spins connected to the loop via the 
dangling C bonds. 
}
\end{figure}

We define first the separable Hilbert space $\mathcal{H}$ of the quantum 
problem as the span of the preferred, orthonormal basis states 
$|\mathcal{C}\rangle$, 
labeled by the configurations $\mathcal{C}\in\mathcal{S}$ of the 
three-coloring model. 
Since we are given the classical energy $E^{\ }_{\mathcal{C}}$ 
of any such configuration $\mathcal{C}$, we only need to construct the 
quantum kinetic energy to complete the definition of a quantum 
Hamiltonian that admits an SMF representation 
with the three-coloring model emerging from its GS. 
To this end, we note that the only classical updates 
allowed by the constraint involve exchanging the colors along closed, 
non-self-intersecting loops given by the sequence of two alternating colors 
starting at some site $i$. In the spin language, any such update amounts to 
nothing but flipping all the spins along the loop, 
i.e., it can be thought of as being induced perturbatively by 
a transverse Ising field 
$\widehat{H}^{\ }_{\textrm{pert.}}\propto\sum_{i}\hat{\sigma}^{\mathrm{x}}_i$. 
Next, we need to construct the set $\mathcal{L}^r=\{\ell\}$ 
of appropriately decorated loops so that: 
(i)~they label all possible loop updates, and 
(ii)~they carry enough information for the energy 
difference $E^{\ }_{\mathcal{C}'}-E^{\ }_{\mathcal{C}}$ 
between two resonating configurations $\mathcal{C}$ and $\mathcal{C}'$ 
to be written as a function $\varepsilon^{\ }_{\ell}$. 
Both requirements are met when a \textit{decorated} loop stands for: 
(1)~The backbone structure of the loop, i.e., the sequence of sites visited 
by the loop. 
(2)~The two-color sequence covering the loop. 
(3)~The values of the spins that do not belong to the loop, 
but are connected to it by a single (dangling) bond. 
An example of such \textit{decorated} loops is given in 
Fig.~\ref{fig: 3col decor}. 
So far, the three-coloring model has been equipped with 
classical stochastics through Monte Carlo simulations using the Metropolis 
algorithm~\cite{Castelnovo04}. 
Correspondingly, we choose to implement the classical stochastics
through a Master equation with 
transition matrix of the Metropolis form. 
We are now in the position to write the SMF decomposition of the quantum 
Hamiltonian whose GS normalization is nothing but 
the classical partition function of the classical three-coloring model, 
\begin{equation}
\label{eq: RK H 3-col}
\widehat{H}^{\ }_{SMF} = \frac{1}{2} \sum_{\ell\in\mathcal{L}^r} 
  w^{\ }_{\ell} 
  \left[ 
    e^{-K \varepsilon^{\ }_{\ell}/2} \widehat{P}^{\ }_{\ell} + 
    e^{+K \varepsilon^{\ }_{\ell}/2} \widehat{P}^{\ }_{\overline{\ell}} 
  - \left( 
      \widehat{T}^{\ }_{\ell,\overline{\ell}} + 
      \widehat{T}^{\ }_{\overline{\ell},\ell} 
    \right)
  \right]. 
\end{equation}
The rescaling coefficient $K\in\mathbb{R}$ can be chosen at will. 
The loop energy $\varepsilon^{\ }_{\ell}$ is given by the integrability 
condition~(\ref{eq: integrability conds a}): 
$\varepsilon^{\ }_{\ell}=E^{\ }_{\mathcal{C}'}-E^{\ }_{\mathcal{C}}$. 
The positive coefficient $w^{\ }_{\ell}$ is chosen to obey the 
Metropolis condition~(\ref{Metrop_w}). 
The operators 
$\widehat{P}^{\ }_{\ell}$ 
and 
$\widehat{T}^{\ }_{\ell,\overline{\ell}}$ 
are defined by their action on the basis states 
$|\mathcal{C}\rangle$~\cite{note: ell in C} 
\begin{equation}
\widehat{P}^{\ }_{\ell} \, |\mathcal{C}\rangle = \left\{ 
  \begin{array}{ll}
    |\mathcal{C}\rangle, & \qquad \textrm{if } \ell\in\mathcal{C}, \\
    &\\ 
    0,                   & \qquad \textrm{otherwise},
  \end{array}
\right.  
\end{equation}
and 
\begin{equation}
\widehat{T}^{\ }_{\ell,\overline{\ell}} \, |\mathcal{C}\rangle:= \left\{ 
  \begin{array}{ll}
    |\overline{\mathcal{C}}\rangle, 
& 
\qquad \textrm{if }\ell\in\mathcal{C}, 
\\
    &\\ 
    0,                              
& 
\qquad \textrm{otherwise}. 
  \end{array}
\right. 
\end{equation}
As it should be, the GS wavefunction in the preferred basis of the Hilbert 
space is 
\begin{equation}
|\Psi^{\ }_{SMF}\rangle = 
  \frac{1}{\sqrt{Z(K)}} 
  \sum_{\mathcal{C}\in\mathcal{S}} 
    e^{-KE^{\ }_{\mathcal{C}}/2}
    |\mathcal{C}\rangle. 
\end{equation}
It is argued in Ref.~\cite{Castelnovo05} that a signature of 
dynamical glassiness 
can indeed be found in this quantum system when it is brought out 
of equilibrium through the local coupling to a heat bath. 

We shall now refine what is meant above by locality. At the classical level in 
thermal equilibrium defined by Eqs.~(\ref{def: 3color model}), 
locality is implemented by a pairwise interaction between the Ising spins that 
is short-ranged. Changing the range of the interaction leaves the backbone of 
a decorated loop $\ell\in\mathcal{L}^{r}$ unchanged while it does change its 
decoration. 
For example, if we modify~(\ref{eq: E_c}) by allowing a next-nearest neighbor 
Heisenberg interaction, we must then decorate a backbone with first and second 
nearest-neighbor dangling bonds. 
At the classical level out of equilibrium, i.e., at the quantum level, 
locality is implemented by the choice of the dependence on $\ell$ of 
$w^{\ }_{\ell}$. 
Since loops of all sizes participate to the classical stochastics 
(quantum dynamics) and 
since loops involve a number of Ising spins proportional to their perimeter, 
large loop updates (tunneling processes) must be penalized compared to small 
loop updates. Stochastic (dynamical) locality is thus achieved when 
$w^{\ }_{\ell}$ decreases with the loop perimeter in an exponential fashion, 
say. 
We conclude that, in the Ising spin language, the (weak) coupling to the bath 
is local if it is between the $\hat{\sigma}^{\mathrm{x}}_{i}$ and a bath of 
harmonic oscillators. 

We close this discussion of the quantum three-coloring model that admits
an SMF decomposition by illustrating how it can be thought of as a 
special point in parameter space of a more general quantum Hamiltonian, 
in the same preferred basis. 
Start from the quantum Ising model in a transverse field 
\begin{equation}
\widehat{H}^{\ }_{Ising} = 
- J \sum_{\langle i,j \rangle} 
  \hat{\sigma}^{\mathrm{z}}_i \hat{\sigma}^{\mathrm{z}}_j
+
\Gamma\sum_{i} \hat{\sigma}^{\mathrm{x}}_i 
+ 
U\sum_{\mathrm{hex}} 
\left[
\cos\left(2\pi\sum_{i\in\mathrm{hex}}\hat{\sigma}^{\mathrm{z}}_{i}/3\right)-1
\right]^2
\label{eq: constrained quantum Ising}
\end{equation}
defined on the honeycomb lattice. The large $U$ limit (i.e., 
$U/J,\;U/\Gamma\to\infty$) restricts the physical Hilbert space to 
the one obtained by imposing on every elementary plaquette (hexagon) 
of the honeycomb lattice the hard constraint 
\begin{equation}
\cos\left(2\pi\sum_{i\in\mathrm{hex}}\hat{\sigma}^{\mathrm{z}}_{i}/3\right)=1.
\end{equation}
In the same way as the $t-J$ model follows from the Hubbard model in the 
large $U$ limit, the quantum three-coloring Hamiltonian 
\begin{equation}
\label{eq: H 3-col}
\widehat{H}^{\ }_{\mathrm{3-c}} = \frac{1}{2} \sum_{\ell\in\mathcal{L}^r} 
  \left[
    \alpha^{\ }_{\ell}\widehat{P}^{\ }_{\ell}
  +
    \alpha^{\ }_{\overline{\ell}}\widehat{P}^{\ }_{\overline{\ell}}
  - 
    \gamma^{\ }_{\ell}
  \left(
  \widehat{T}^{\ }_{\ell,\overline{\ell}}
  +
  \widehat{T}^{\ }_{\overline{\ell},\ell}
  \right)
  \right]
\end{equation}
follows from~(\ref{eq: constrained quantum Ising}) 
when choosing the data $(\alpha^{\ }_{\ell},\gamma^{\ }_{\ell})$ 
properly~\cite{Castelnovo05}. The Hamiltonian~(\ref{eq: RK H 3-col}) 
is another point in the parameter space 
$(\alpha^{\ }_{\ell},\gamma^{\ }_{\ell})\in\mathbb{R}^2$ 
specified by conditions~(\ref{eq: condition for positive definite}) 
and conditions~(\ref{eq: integrability conds}). 

As a second example, we start from the medial lattice of the honeycomb 
lattice. This is the Kagome lattice obtained from the mid-points of the 
nearest-neighbor bonds of the honeycomb lattice. If we 
attach to any site of the Kagome lattice a quantum spin carrying 
angular momentum one, this gives a three-dimensional Hilbert 
space per site, and a global Hilbert space $\mathcal{H}$ of dimension 
$3^{N}$, with $N$ the total number of sites on the Kagome lattice. 
The three quantum numbers allowed for the component of the spin along the 
quantization (z) axis can be identified with the three colors $A$, $B$, and 
$C$ of the quantum three-coloring model. Next, we impose two constraints. 
First, we demand that the coarse-grained magnetization along the quantization 
axis vanishes. Here, the coarse-grained magnetization is defined locally on 
the honeycomb lattice by attaching to any site of the honeycomb lattice 
the arithmetic average over the spin quantum numbers along the quantization 
axis defined on the bonds meeting at this site. Second, we remove from 
$\mathcal{H}$ any state that has all three spins with vanishing projection 
along the quantization axis when these three spins define a vertex of the 
honeycomb lattice. The combination of such two conditions on $\mathcal{H}$ 
is equivalent to imposing the three-color constraint. 
Having defined the physical Hilbert space  $\mathcal{H}^{\ }_{\mathrm{phys}}$, 
we can then proceed as with the quantum Ising spin in a transverse magnetic 
field and identify a quantum Hamiltonian with nearest-neighbor spin-one 
interaction, say, with Hamiltonian~(\ref{eq: H 3-col}). Closely related 
quantum spin-one models can be found in Ref.~\cite{Wen03}. 

\section{Conclusions}
\label{sec: Conclusions}

The study of quantum Hamiltonians that are of the Rokhsar-Kivelson type 
when represented in a preferred basis has proven fruitful in the study of 
quantum critical points. 
This is so because of an intimate relation between the matrix 
representations of these quantum Hamiltonians in preferred bases 
and some combinatorial problems of classical statistical physics. 

We have shown in this paper that the examples of quantum Hamiltonians
encountered this far, when fine-tuned to their RK points, 
belong to a broader class of real, symmetric, and irreducible matrices 
that are in one-to-one correspondence with matrices known in the mathematic 
literature as stochastic matrices. 
Any stochastic matrix can be used to define a Master equation of the matrix 
type that encodes the approach to thermal equilibrium of a classical 
system coupled to a heat bath. 
It is then natural to generalize the known examples of quantum Hamiltonians 
that are represented by matrices obeying the RK conditions (I-III)
to any quantum Hamiltonian 
that admits a matrix representation in this broader class, that we dub the 
\textit{Stochastic Matrix Form} (SMF) decomposition. 
Thus, some (not necessarily all) quantum phase transitions 
induced by tuning the coupling constants, 
say $K$, $\mu$, $B$, etc., of a quantum Hamiltonian represented by a matrix 
which is SMF decomposable can be understood as classical phase transitions 
obtained by tuning the reduced inverse temperature $K$, the chemical 
potential $\mu$, the external field $B$, etc., that enter a classical 
partition function as intensive thermal variables. 
This correspondence between the quantum GS and a classical partition function 
extends to the excitations of the quantum system and the relaxation modes of 
the classical system when the latter is properly coupled to a thermal bath. 

The correspondence between a quantum Hamiltonian represented by a matrix 
which is SMF decomposable and Master equations of the Matrix type 
opens the intriguing possibility to find a signature in a quantum system 
for exotic properties of a stochastic classical system such as aging and 
dynamical glassiness. From a more practical point of view, this also 
suggests a classical Monte Carlo alternative to a quantum Monte Carlo 
simulation when probing numerically the excitation spectrum of a quantum 
system that allows for an SMF decomposition. 

Finally, we have also proven that any quantum Hamiltonian, up to a shift of 
its eigenvalues, admits a continuous manifold of bases in which its 
representation is SMF decomposable. In this sense, the correspondence 
between a quantum Hamiltonian construed as an abstract 
operator acting on some separable Hilbert space and the stochastic dynamics 
describing the relaxation to thermal equilibrium of discrete classical 
statistical systems coupled to heat baths is one-to-continuously many. 

We are indebted to Christopher Henley for a critical reading of our manuscript 
and for challenging us to explore the behavior of the SMF decomposition 
under a basis transformation.

\end{document}

%% file: tridimer1.tex
\setlength{\unitlength}{0.06mm}
\begin{picture}(90,50)(0,10) 
  \linethickness{0.5mm}
  \put(30,5){\line(1,0){50}} 
  \linethickness{0.003mm}
  \put(30,5){\circle*{10}}
  \multiput(29,5)(-0.56,0.866){44}{\line(1,0){0.1}}
  \put(80,5){\circle*{10}}
  \linethickness{0.5mm}
  \put(5,43){\line(1,0){50}}
  \linethickness{0.003mm}
  \put(5,43){\circle*{10}}
  \multiput(55,43)(0.56,-0.866){44}{\line(1,0){0.1}}
  \put(55,43){\circle*{10}}
  \multiput(54,43)(-0.56,-0.866){44}{\line(1,0){0.1}}
\end{picture}

%% file: tridimer2.tex
\setlength{\unitlength}{0.06mm}
\begin{picture}(90,50)(0,10) 
  \linethickness{0.03mm}
  \put(30,5){\line(1,0){50}} 
  \put(30,5){\circle*{10}}
  \multiput(26,5)(-0.56,0.866){44}{\line(1,0){8}}
  \put(80,5){\circle*{10}}
  \put(5,43){\line(1,0){50}}
  \put(5,43){\circle*{10}}
  \multiput(51,43)(0.56,-0.866){44}{\line(1,0){8}}
  \put(55,43){\circle*{10}}
  \multiput(55,43)(-0.56,-0.866){44}{\line(1,0){0.1}}
\end{picture}

%% file: tridimerplaq.tex
\setlength{\unitlength}{0.06mm}
\begin{picture}(90,50)(0,10) 
  \linethickness{0.03mm}
  \put(30,5){\line(1,0){50}} 
  \put(30,5){\circle*{10}}
  \linethickness{0.003mm}
  \multiput(29,5)(-0.56,0.866){44}{\line(1,0){0.1}}
  \linethickness{0.03mm}
  \put(80,5){\circle*{10}}
  \put(5,43){\line(1,0){50}}
  \put(5,43){\circle*{10}}
  \linethickness{0.003mm}
  \multiput(55,43)(0.56,-0.866){44}{\line(1,0){0.1}}
  \put(55,43){\circle*{10}}
  \multiput(54,43)(-0.56,-0.866){44}{\line(1,0){0.1}}
\end{picture}

%% file: tridimer1_sm.tex
\setlength{\unitlength}{0.025mm}
\begin{picture}(90,50)(0,10) 
  \linethickness{0.208mm}
  \put(30,5){\line(1,0){50}} 
  \linethickness{0.001mm}
  \put(30,5){\circle*{10}}
  \multiput(29,5)(-0.56,0.866){44}{\line(1,0){0.1}}
  \put(80,5){\circle*{10}}
  \linethickness{0.208mm}
  \put(5,43){\line(1,0){50}}
  \linethickness{0.001mm}
  \put(5,43){\circle*{10}}
  \multiput(55,43)(0.56,-0.866){44}{\line(1,0){0.1}}
  \put(55,43){\circle*{10}}
  \multiput(54,43)(-0.56,-0.866){44}{\line(1,0){0.1}}
\end{picture}

%% file: tridimer2_sm.tex
\setlength{\unitlength}{0.025mm}
\begin{picture}(90,50)(0,10) 
  \linethickness{0.001mm}
  \put(30,5){\line(1,0){50}} 
  \put(30,5){\circle*{10}}
  \multiput(26,5)(-0.56,0.866){44}{\line(1,0){8}}
  \put(80,5){\circle*{10}}
  \put(5,43){\line(1,0){50}}
  \put(5,43){\circle*{10}}
  \multiput(51,43)(0.56,-0.866){44}{\line(1,0){8}}
  \put(55,43){\circle*{10}}
  \multiput(55,43)(-0.56,-0.866){44}{\line(1,0){0.1}}
\end{picture}

%% file: paper_05_01_25.bbl
\begin{thebibliography}{99}
\bibitem{note: LS}
The assumption that the Hilbert space is fully connected under the action 
of the time-evolution operator implies that the Hamiltonian matrix in the 
Hilbert space basis is irreducible. 
This requires in turn that no row or column in the matrix has less than one 
non-vanishing off-diagonal element. Since the Hamiltonian is symmetric, we 
immediately obtain a lower bound for $|\mathcal{L}|$, namely 
$|\mathcal{L}| \geq 2(|\mathcal{S}|-1)$, and Eq.~(\ref{eq: def H}) holds.
\bibitem{note: 2x2 RK matrix}
Notice that the operator $\widehat{Q}^{\ }_{\mathcal{C},\mathcal{C}'}$ acts on 
the two-dimensional subspace of the Hilbert space given by the span of the 
two basis states $|\mathcal{C}\rangle$ and $|\mathcal{C}'\rangle$. Hence,
$\widehat{Q}^{\ }_{\mathcal{C},\mathcal{C}'}$ can be represented by the
$2\times2$ matrix
%
%
\begin{equation*}
\left( 
  \begin{array}{cc} 
  e^{-K (E^{\ }_{\mathcal{C}'}-E^{\ }_{\mathcal{C}})/2} 
  & 
  -1 \\ 
  -1 
  & 
  e^{+K (E^{\ }_{\mathcal{C}'}-E^{\ }_{\mathcal{C}})/2}
  \end{array}
\right). 
\end{equation*}
\bibitem{note: ell in C} 
For conciseness, the symbolic notation 
$\in\mathcal{C}$ ($\ell\in\mathcal{C}$) 
means that the plaquette covering 
$$ (decorated loop $\ell$)  
appears in configuration $\mathcal{C}$.
\bibitem{note: RK dimer conditions} 
Provided we satisfy the conditions on the parameters 
$w^{\ }_{}$, $K$, and 
$\varepsilon^{\ }_{}$, which are: 
$w^{\ }_{} = w^{\ }_{} > 0$, 
$K \geq 0$, and 
$\varepsilon^{\ }_{}$ integrable.
\bibitem{note: q8v integrability}
Notice that the simple additive nature of the energy in the eight-vertex 
model, where the Boltzmann weights are given by the product of the fugacities 
for the vertices, allows to satisfy trivially the integrability conditions on 
$\varepsilon^{\ }_{\ell}$.
\end{thebibliography}
